\documentclass[%
 reprint,
superscriptaddress,
 amsmath,amssymb,
 aps,%prl
prb,
]{revtex4-2}

\usepackage{graphicx}% Include figure files
\usepackage{dcolumn}% Align table columns on decimal point
\usepackage{bm}% bold math
\usepackage[overload]{empheq} % subequations
\usepackage[hidelinks]{hyperref}% add hypertext capabilities
\usepackage{xcolor}
\hypersetup{breaklinks=true}
\newcommand{\rme}{{\rm e}}
\newcommand{\rmd}{{\rm d}}
\newcommand{\rmi}{{\rm i}}
\newcommand{\rmF}{{\rm F}}
\newcommand{\bk}{{\boldsymbol{k}}}

\newcommand{\bx}{{\boldsymbol{x}}}
\newcommand{\bX}{{\boldsymbol{X}}}
\newcommand{\bY}{{\boldsymbol{Y}}}
\newcommand{\bu}{{\boldsymbol{u}}}
\newcommand{\nabX}{\boldsymbol{\nabla}_{\!\!\boldsymbol X}}
\newcommand{\nabY}{\boldsymbol{\nabla}_{\!\!\boldsymbol Y}}
\newcommand{\nabx}{\boldsymbol{\nabla}_{\!\!\boldsymbol x}}
\newcommand{\nabk}{\boldsymbol{\nabla}_{\!\!\boldsymbol k}}

\usepackage{dsfont}

\def\Eq#1{\hyperref[#1]{Eq.~(\ref*{#1})}}
\def\Eqs#1{\hyperref[#1]{Eqs.~(\ref*{#1})}}

\begin{document}

%\preprint{APS/123-QED}

\title{Kinetics of information scrambling in correlated metals: \texorpdfstring{\\}{} disorder-driven transition from shock-wave to FKPP dynamics}

\author{Camille Aron}
 \email{aron@ens.fr}
 \affiliation{Laboratoire de Physique de l'\'Ecole Normale Sup\'erieure, ENS, Universit\'e PSL, 
 CNRS, Sorbonne Universit\'e, Universit\'e Paris Cit\'e, F-75005 Paris, France}
 \affiliation{Institute of Physics, \'Ecole Polytechnique F\'ed\'erale de Lausanne (EPFL), CH-1015 Lausanne, Switzerland}
 \author{\'Eric Brunet}
 \affiliation{Laboratoire de Physique de l'\'Ecole Normale Sup\'erieure, ENS, Universit\'e PSL, 
 CNRS, Sorbonne Universit\'e, Universit\'e Paris Cit\'e, F-75005 Paris, France}
\author{Aditi Mitra}
\affiliation{Center for Quantum Phenomena, Department of Physics, New York University, 726 Broadway, New York, NY, 10003, USA}

%\date{\today}% It is always \today, today,
             %  but any date may be explicitly specified

\begin{abstract}
Quenched disorder slows down the scrambling of quantum information. Using a bottom-up approach, we formulate a kinetic theory of scrambling in a correlated metal near a superconducting transition, following the scrambling dynamics as the impurity scattering rate is increased. Within this framework, we rigorously show that the butterfly velocity $v$ is bounded by the light cone velocity $v_{\rm lc }$ set by the Fermi velocity. We analytically identify a disorder-driven dynamical transition occurring at small but finite disorder strength between a spreading of information characterized at late times by a discontinuous shock wave propagating at the maximum velocity $v_{\rm lc}$, and a smooth traveling wave belonging to the Fisher or Kolmogorov-Petrovsky-Piskunov (FKPP) class and propagating at a slower, if not considerably slower, velocity $v$. In the diffusive regime, we establish the relation $v^2/\lambda_{\rm FKPP} \sim D_{\rm el}$ where $\lambda_{\rm FKPP}$ is the Lyapunov exponent set by the inelastic scattering rate and $D_{\rm el}$ is the elastic diffusion constant.
\end{abstract}

%\keywords{Suggested keywords}%Use showkeys class option if keyword
                              %display desired
\maketitle

Information scrambling refers to the efficient spreading and loss of information throughout an extended many-body system.
Its characterization is fundamental to the foundations of quantum chaos and has implications in the development quantum computing technologies.
The picture that has emerged from the exact or numerical solutions to a variety of classical and quantum thermalizing models with local interactions is 
a ballistic spreading of information at a so-called butterfly velocity.
In dual-unitary circuits, the information scrambling occurs precisely on the light rays propagating at the maximum velocity allowed by causality~\cite{Prosen2019,Lamacraft2020}.
In random quantum circuits made of Haar-distributed unitaries, 
scrambling dynamics was related to classical growth processes with slower ballistic fronts that broaden either diffusively in $d=1$, or according to fluctuations governed by the Kardar-Parisi-Zhang universality class in $d=2$ and $d=3$~\cite{Nahum2018,Piroli2020}.

Alongside these minimal models, as well as classical~\cite{Moessner2018a,AdamHuse2018,Banerjee2021,ManasHuse2023}, semi-classical, large $N$ or holographic models~\cite{Stanford2017,Swingle2017_ON,Shenker2014,Swingle2016}, it is essential to address those questions in realistic situations where exact or numerical solutions are out of reach.
Aleiner, Faoro and Ioffe articulated those questions in the larger framework of electronic transport by deriving a quantum kinetic equation for out-of-time-ordered correlators (OTOCs) within a so-called many-world Keldysh formalism~\cite{Aleiner2016}.
They argued that the scrambling dynamics in metals with either phonon or Coulomb interactions are governed by Fisher or Kolmogorov-Petrovsky-Piskunov (FKPP) equations, resulting in smooth non-broadening scrambling fronts propagating at a butterfly velocity set by the Fermi velocity, $v_\rmF$. Interestingly, it was argued that the presence of disorder would significantly reduce this butterfly velocity~\cite{Aleiner2016,SwingleSachdev2017,Swingle2017_disorder,Galistski2018}.

Building on the approach of Ref.~\cite{Aleiner2016}, we worked out the kinetics of quantum information scrambling in a paradigmatic model of clean interacting metals in the vicinity of a superconducting phase transition, where electron-electron interactions are dominated by superconducting fluctuations~\cite{ABM}.
We found scrambling fronts that travel at $v_\rmF$ but do not belong to the FKPP universality class. Remarkably, their late-time spatial profiles develop a shock-wave discontinuity at the boundary of the light cone.

\begin{figure}
    \centering
    \input{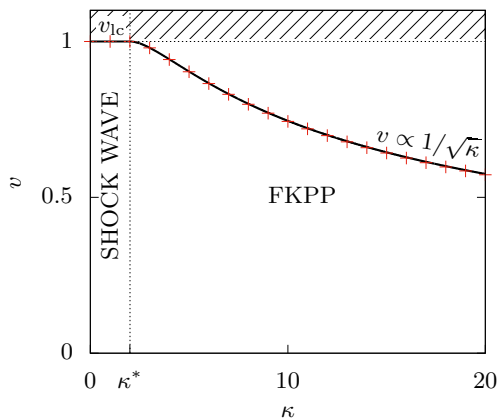}
    \caption{Quantum butterfly velocity $v$ as a function of the disorder strength $\kappa$. For $\kappa < \kappa^*$, the front propagates at the light cone velocity $v_{\rm lc} = 1$ in units of $v_\rmF / \sqrt{d}$. For $\kappa > \kappa^*$, the solid line is the FKPP prediction made in Eq.~(\ref{eq:v_FKPP}). The red marks are numerical results obtained by solving Eqs.~(\ref{eq:coupled_PDEs_adim}) for $d=1$ up to times $\tau = 3000$
    ($\gamma=1$ \textit{i.e.}\@ $\kappa^*:=1+\gamma=2$).
    }    \label{fig:velocity}
\end{figure}

In this Letter, we investigate the impact of impurity scattering on the dynamics of the scrambling front. Momentum relaxation is indeed significant in metals with an elastic timescale typically in the tens of femtoseconds, \textit{i.e.}\@ much shorter than the scattering time due to electronic interactions, especially at low temperatures. 
We first derive an effective set of two coupled partial differential equations (PDEs) governing the scrambling dynamics at large scales. Then, we analytically elucidate how the traveling-wave solutions that develop at late times are affected by the presence of disorder, from the clean case to the diffusive regime.
We find that the shock-wave phenomenology found in the clean case is robust against weak disorder. However, we unravel a dynamical phase transition that causes the scrambling kinetics to abruptly conform to the FKPP class when the disorder strength exceeds a critical value, see Fig.~\ref{fig:velocity}. We fully characterize the information scrambling in the FKPP regime by working out the profiles of the traveling fronts and their butterfly velocities. 

\paragraph*{Model.}
We consider a system of interacting electrons in $d\geq2$ dimensions that are subject to both elastic and inelastic scattering. The former is due to static non-magnetic impurities and defects. The latter is due to electron-electron interactions in the Cooper channel. This choice is guided by the relatively straightforward treatment of superconducting fluctuations within the random phase approximation. However, our approach can be extended to other models with different interactions as long as a quasi-particle description is valid.
For concreteness, we have in mind the Hamiltonian
\begin{align} \label{eq:BCS}
    H = & \sum_{\bk,\, \sigma} \epsilon_\bk  c^\dagger_{\bk\sigma} c_{\bk\sigma} + \sum_{\bk\bk', \, \sigma} V_{\bk-\bk'} c^\dagger_{\bk\sigma} c_{\bk'\sigma}  \nonumber \\
    & +   U \!\!\! \sum_{\bk\bk'\bk'', \, \sigma} \!\! c^\dagger_{\bk\sigma}c^\dagger_{-\bk+\bk'\bar\sigma}c_{\bk''+\bk'\bar\sigma}c_{-\bk''\sigma}.
\end{align}
The operator $c_{\bk\sigma}^\dagger$ creates a fermion with  spin $\sigma = \uparrow$  or $\downarrow$ ($\bar\sigma = \downarrow$, $\uparrow$) and momentum $\bk$ in the Brillouin zone. $\epsilon_\bk$ is the dispersion relation.
Electronic energies are measured relative to the chemical potential and $E_\rmF$ is the Fermi energy. 
For simplicity, we shall assume a spherical Fermi surface, \textit{i.e.}\@ $\epsilon_\bk = 0$ when $k \to k_\rmF$.

The attractive interaction $U<0$ facilitates superconductivity.
In dimensions $d>2$, this model exhibits a finite-temperature phase transition towards a superconducting phase associated with the spontaneous breaking of the $U(1)$ symmetry. In $d=2$, it is replaced by a Berezinskii–Kosterlitz–Thouless (BKT) transition with quasi-long-range order~\cite{Kosterlitz1973}.
Here, we work near criticality in the normal phase of the BCS-type superconductor and in the regime of validity of the Ginzburg-Levanyuk criterion, where the superconducting fluctuations are sizable but weakly interacting~\cite{Larkin2002,Lemonik2018b}.

The disordered potential $V(\bx)$ is assumed to be short-ranged and Gaussian-distributed, with covariance $\langle V(\bx) V(\bx')\rangle  - \langle V(\bx) \rangle^2 = g \delta(\bx-\bx')$ with $g>0$~\footnote{More generally, we can assume that the higher-order cumulants of the disordered potential are irrelevant in an renormalization group sense.}.
We work far from localization regimes, where the disorder can be treated in a classical fashion, \textit{i.e.}\@ not accounting for coherent effects between scattering trajectories.
In practice, the impurity scattering is treated with the Born approximation to second order in $g$~\footnote{The terms at order $g$ can be absorbed by a shift of the dispersion relation.}.

We formulate the kinetic theory of quantum information scrambling by starting from the quantum kinetic equation on the many-world distribution functions $F_{\alpha\beta}(t,\bx;\omega,\bk)$, where the indices $\alpha, \beta \in \{ u,d \}$ span two replicated worlds (up and down)~\cite{Aleiner2016,ABM}.
The intraworld components $\alpha  = \beta$ correspond to the standard electronic distribution functions. We concentrate on the interworld components $F_{\alpha  \neq \beta}$ which are directly related to four-point OTOCs and to the growth of operators. In the gradient approximation, valid when the microscopic scales set by $\hbar/E_\rmF$ and $1/k_\rmF$ are much shorter than the spatiotemporal variations of $F_{\alpha\beta}$, the dynamics of the latter are governed by a non-linear partial-integrodifferential equation reading
\begin{align} \label{eq:QKineticEq}
    [\partial_t + \boldsymbol{v}_\bk \cdot \nabx] F_{\alpha\beta} = I_{\alpha\beta},
\end{align}
with the velocity $\boldsymbol{v}_\bk := \nabk \epsilon_\bk$ and where the collision integral $I_{\alpha\beta}$ is a non-linear functional of the $F_{\alpha\beta}$'s collecting contributions from the disordered potential and the electronic interactions. 
The kinetic theory is considerably simplified by working (i) in terms of the first two components of a partial-wave expansion in $\bk$, (ii) on-shell, \textit{i.e.} $\omega\to\epsilon_\bk$, and (iii) near the Fermi surface, \textit{i.e.}  $k \to k_\rmF$. This amounts to working with the ansatz
\begin{align}
    F_{\alpha\beta}(t,\bx;\omega,\bk) \mapsto \epsilon_{\alpha\beta} \left[ \phi(t,\bx)  + \bu_\bk \cdot \boldsymbol{\phi}_1(t,\bx)  \right] ,
\end{align}
with $\epsilon_{du} = - \epsilon_{ud} = 1$,the unit vector $\bu_\bk : = \bk/k$, and where $\phi$ and $\boldsymbol{\phi}_1$ 
are the isotropic and first anisotropic corrections, respectively, to the on-shell interworld distribution function. 
The validity of the above ansatz is discussed for the clean case in Ref.~\cite{ABM}, and the partial-wave truncation is all the more accurate in the presence of impurity scattering as it reduces momentum anisotropy.
Following the steps detailed in~\autoref{app:model}, the kinetic equation can be brought to a set of coupled non-linear PDEs reading 
\begin{align}
\left\{     \label{eq:coupled_PDEs}
\begin{array}{rl}  
     \partial_t \phi +  \frac{v_\rmF}{d} \nabx   \cdot \boldsymbol{\phi}_1 \!\!\!\!& =  \phi (\phi^2-1)/\tau_{\rm sc}
 \\[\medskipamount]
          \partial_t  \boldsymbol{\phi}_1 +  v_\rmF \nabx \phi  \!\!\!\!& = \boldsymbol{\phi}_1  (\gamma\phi^2-1) /\tau_{\rm sc} - \boldsymbol{\phi}_1 / \tau_{\rm el}.
\end{array}
\right.
\end{align}
This represents an effective kinetic theory of scrambling in terms of the two fields $\phi(t,\bx)$ and $\boldsymbol{\phi}_1(t,\bx)$.
The dimensionless parameter $\gamma$ tunes the distance to the superconducting transition: $\gamma=1$ corresponds to criticality and $0 < \gamma < 1$ to off-critical regimes in the normal phase. 
$\tau_{\rm el} \propto 1/g$ is the elastic timescale due to scattering on the disordered potential and $\tau_{\rm sc}$ is the timescale set by the inelastic scattering on the superconducting fluctuations (Cooperons). 
These parameters of the model~(\ref{eq:coupled_PDEs})
depend in a non-trivial fashion on those of the original microscopic model~(\ref{eq:BCS}) and have to be understood as renormalized quantities resulting from a complex cross-feed. For example, disorder is known to enhance the inelastic scattering rate $1/\tau_{\rm sc}$ at low temperatures~\cite{ALTSHULER1985}.
The PDEs~(\ref{eq:coupled_PDEs}) have a ``correlated-world'' solution, $\phi = 1$ and $\boldsymbol{\phi}_1 = 0$, corresponding to both replicated worlds evolving coherently and is expected to be unstable for chaotic systems. $\phi = \boldsymbol{\phi}_1 = 0$ is the ``uncorrelated-world'' solution, corresponding to a total loss of coherence between worlds. 

At $\tau_{\rm el} \to \infty$, we recover the clean model studied in Ref.~\cite{ABM}.
In the non-interacting limit, $\tau_{\rm sc} \to \infty$, and in the diffusive regime (at late times) where $\partial_t \boldsymbol{\phi}_1 $ can be neglected relative to $\boldsymbol{\phi}_1/\tau_{\rm el}$, the above coupled PDEs reduce to a simple diffusion equation: $\partial_t \phi - D_{\rm el} \nabx^2 \phi = 0$ with the elastic diffusion coefficient $D_{\rm el} := {v_\rmF^2 \tau_{\rm el}}/d$. 
In this Drude limit, the correlated-world solution $\phi = 1$ is stable against local perturbations, expressing the absence of quantum information scrambling when only disorder is present.
In the generic case with both elastic and inelastic scattering, we study how the scrambling dynamics evolve as 
the dimensionless disorder coupling constant $\kappa := \tau_{\rm sc}/ \tau_{\rm el}\geq 0$ is increased from the clean to the diffusive metal~\footnote{In particular, we do not start from the Usadel equation as it relies on a diffusive approximation, see Ref.~\cite{kamenev2023}}.
The analysis is simplified by rescaling time and space,
$ \tau := t / \tau_{\rm sc} \mbox{ and } \bX := \bx / \ell_{\rm sc}$ with $\ell_{\rm sc} := v_\rmF \tau_{\rm sc} / \sqrt{d}$, together with $\boldsymbol{\phi}_1 \mapsto \sqrt{d}\, \boldsymbol{\phi}_1$.
The above PDEs become
\begin{align}
\left\{     \label{eq:coupled_PDEs_adim}
\begin{array}{rl}  
     \partial_\tau \phi +  \nabX \cdot  \boldsymbol{\phi}_1 \!\!\!\!& =  \phi (\phi^2-1)
 \\ [\medskipamount]
          \partial_\tau  \boldsymbol{\phi}_1 +  \nabX \phi  \!\!\!\!& = \boldsymbol{\phi}_1  ( \gamma\phi^2-1) - \kappa \boldsymbol{\phi}_1.
\end{array}
\right.
\end{align}
The initial conditions are taken as local and spherically symmetric perturbations to the correlated-world solution:
\begin{align}
\left\{
\begin{array}{rl}
     \phi(\tau=0,\bX) &=  1 - \delta\phi_0(X)\\
    \boldsymbol{\phi}_1(\tau=0,\bX) &=  0,
\end{array}
    \right.
\end{align}
with the radial coordinate $X := \|\bX\|$  and a perturbation $0 \leq \delta\phi_0(X) \ll 1 $ which is non-vanishing on small support of radius $R_0$ and $\delta\phi_0(X > R_0) = 0$. This guarantees that subsequent dynamics reduce to an effective one-dimensional problem for $\phi(\tau,X)$ and $\phi_1(\tau,X)$ along the radial direction. 
The late-time dynamics marginally depend on the choice of $\delta\phi_0(X)$.

The instability of the correlated-world solution against local perturbations is expected to generate a transient state where both solutions ($\phi\simeq 1$ and $\phi\simeq0$) are separated by a domain wall, a front, located on a sphere of growing radius. The profile and the motion of this front determine the dynamics of the scrambling of quantum information. 
At late times, the front is located far from the origin and it is governed by the $d=1$ version of the PDEs~(\ref{eq:coupled_PDEs_adim}) where $X$ is now the radial coordinate and $\phi_1$ is the radial component of $\boldsymbol{\phi}_1$~\cite{ABM}.
In~\autoref{app:ballistic}, we provide rigorous proof that this growth is bounded by a maximal velocity set by the Fermi velocity, $v_{\rm lc} = 1$ in units of $ v_{\rm F} / \sqrt{d}$. 
This ensures that the dynamics strictly take place within a causal light cone where $v_{\rm lc}$ acts as the effective speed of light.
This motivates us to look for traveling fronts propagating at a constant velocity $v \leq v_{\rm lc}$ and located at $m_\tau \sim v \tau$ by assuming
\begin{align} \label{eq:traveling_wave}
 \left\{ \begin{array}{rl}
    \phi(\tau,m_\tau  + z  ) &\stackrel{\tau\to\infty}{\longrightarrow} f(z)\\
    \phi_1(\tau, m_\tau +z) &\stackrel{\tau\to\infty}{\longrightarrow} f_1(z).
    \end{array}
    \right.
\end{align}

\paragraph*{Fermi shock-wave dynamics.}
\begin{figure}
    \input{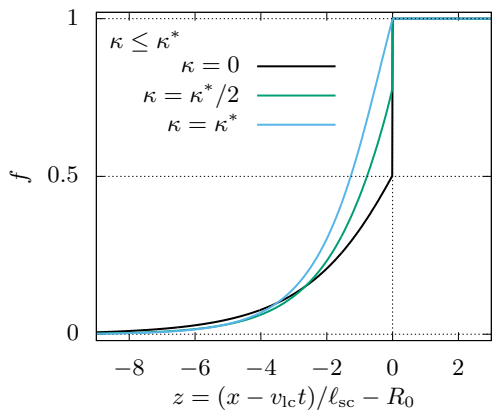}
    \caption{Radial profiles $f(z)$ of the late-time Fermi shock wave traveling at the maximal velocity $v_{\rm lc}$, computed exactly from the coupled PDEs~(\ref{eq:coupled_PDEs_adim}) for $\kappa = 0$, $\kappa^*/2$\, and $\kappa^*$ ($\gamma = 1$ \textit{i.e.}\@ $\kappa^* = 2$.) The offset $R_0$ is the radius of the initial condition.
    The front thickness is of the order of $\ell_{\rm sc}:=v_\rmF \tau_{\rm sc} / \sqrt{d}$ in the original units.}    \label{fig:ballistic}
\end{figure}

In the spirit of Ref.~\cite{ABM}, we first look for traveling-wave solutions propagating at the maximum velocity $v = v_{\rm lc} = 1$ and that are discontinuous at the boundary of the light cone.
We leave the full computation of the front profile to~\autoref{app:ballistic}.
Here, we simply focus on extracting the discontinuity by parametrizing the near-front geometry as
\begin{align*}
 \left\{ \begin{array}{rlrl}
    f(z<0, |z| \ll 1) &\simeq L + z/\xi, & f(z>0) &= 1\\
    f_1(z<0, |z| \ll 1) &\simeq -M - z/\xi_1, & f_1(z>0) &= 0,
    \end{array}
    \right.
\end{align*}
where $L$, $M$, $\xi$, and $\xi_1$ are positive parameters to be determined, and $m_\tau = \tau + R_0$.
We find a critical disorder strength 
\begin{align}
    \kappa^* = 1 + \gamma
\end{align}
separating two distinct solutions.
For $\kappa < \kappa^*$, we find a solution of Eqs.~(\ref{eq:coupled_PDEs_adim}) with a traveling discontinuity from $f(0^-) = L$ to $f(0^+) = 1$, with 
\begin{align}
    L(\kappa < \kappa^*) = \frac{\sqrt{1+4 \kappa^*(1+\kappa)}-1}{2\kappa^*}\label{eq:L}
\end{align}
and $M = 1-L$.
We find finite values of $\xi$, indicating that the thickness of the front is controlled by $\ell_{\rm sc}$.
This generalizes the Fermi shock-wave dynamics identified in Ref.~\cite{ABM} to the weakly-disordered case.
We illustrate this shock-wave profile in~\autoref{fig:ballistic}.
When $\kappa\to\kappa^*$ from below, the discontinuity closes continuously, $L \to 1$, but
the slope remains discontinuous. For $\kappa>\kappa^*$, $L=1$ is the only solution and the finite slope left of the front brutally vanishes, $f'(0) = 0$, signaling the sudden death of the Fermi shock wave. 

\paragraph*{FKKP dynamics.}
Inspired by the FKPP equation proposed in Ref.~\cite{Aleiner2016}, we look for smooth traveling-wave solutions that belong to the FKPP class, with exponential tails ahead of the front of the form
 \begin{align} \label{eq:tails}
  \left\{ \begin{array}{rl}
     f(z \gg 1) &\simeq 1-A z \exp(- \mu z) \\
     f_1(z \gg 1) &\simeq -B z \exp(- \mu z), 
     \end{array}
     \right.
 \end{align}
 where $A$, $B$, and the spatial decay rate $\mu$ are positive parameters. 
To reveal the hidden FKPP nature of the PDEs~(\ref{eq:coupled_PDEs_adim}), we reformulate them in terms of $\delta \phi := 1-\phi$ and $\boldsymbol{\phi}_1$ which are expected to be small and slowly varying ahead of the front, as per Eq.~(\ref{eq:tails}).
Standard algebra detailed in~\autoref{app:FKPP} yields, for any $\kappa \geq 0$,
\begin{equation}  \label{eq:FKPP}
   \left[ \partial_\tau^2   + (\kappa\!-\!\kappa^*) \partial_\tau  - \nabX^2  \right] \! \delta \phi \!=\! 2 (2\!+\!\kappa\!-\!\kappa^*) \, \delta \phi - \mathrm{NL},
\end{equation}
where we collected the non-linear terms under the symbol $\mathrm{NL}$.
The term in $\partial_\tau \delta \phi$ is odd under time reversal and its prefactor changes sign at $\kappa=\kappa^*$. When $\kappa > \kappa^*$, it can be loosely interpreted as a dissipative term, bringing Eq.~(\ref{eq:FKPP}) to a standard FKPP fashion, whereas it acts as a drive when $\kappa < \kappa^*$.
The term $\partial^2_\tau \delta \phi$ can be interpreted as inertia which, to the best of our knowledge, has not been discussed in the broader context of FKPP.
Following a standard FKPP analysis detailed in ~\autoref{app:FKPP_prop},
we find that there are no traveling-wave solutions of the form (\ref{eq:tails}) when $\kappa < \kappa^*$. However, for $\kappa > \kappa^*$, we now find a front propagating at the velocity, in units of $v_\rmF/\sqrt{d}$,
\begin{align}
        v(\kappa>\kappa^*) &= 2 \sqrt{2} \frac{\sqrt{2+\kappa-\kappa^*}}{4+\kappa-\kappa^*} \leq v_{\rm lc}. \label{eq:v_FKPP} 
\end{align}
In the limit $\kappa\to\kappa^*$, we recover $v\to v_{\rm lc}$, consistently with the Fermi shock wave.
$v$ decreases monotonously with increasing disorder strength, and $v \propto 1/\sqrt{\kappa}\to 0$ when $\kappa\to\infty$. 
In \autoref{fig:velocity}, we compare the front velocities extracted from the numerical solutions of the PDEs (\ref{eq:coupled_PDEs_adim}) to the FKPP prediction in Eq.~(\ref{eq:v_FKPP}). The agrement is excellent. 
In~\autoref{app:FKPP_check}, using results of Refs.~\cite{vanSaarloos.2003, Bramson.1978, Roberts.2013, EbertvanSaarloos.2000, BBHR.2016, NolenRoquejoffreRyzhik.2019, BerestyckiBrunetDerrida.2017, BerestyckiBrunetDerrida.2018, Graham.2019, BrunetDerrida.1997}, we provide a meticulous numerical analysis that further demonstrates, beyond any reasonable doubt, the FKPP nature of the front dynamics as soon as $\kappa > \kappa^*$.

Interestingly, the early-time dynamics governed by the linearized version of Eq.~(\ref{eq:FKPP}) are characterized by an exponential growth of the spatially-integrated perturbation $M(\tau) := \int \! \rmd \bX \delta \phi(\tau,\bX) = M(0) \exp(2 \tau)$ where the Lyapunov exponent $\lambda=2$ does not depend on whether the shock-wave or the FKPP regime is governing the scrambling dynamics. In the FKPP regime, the growth of $\delta\phi$ ahead of the front is also exponential, as per Eq.~(\ref{eq:tails}), and the scrambling dynamics can be characterized by a rate $\lambda_{\rm FKPP} = v\mu= 4+8/(\kappa-\kappa^*)$ where $\mu$ is computed in~\autoref{app:FKPP_prop}.

\paragraph*{Diffusive regime.}
\begin{figure}
    \centering
    \input{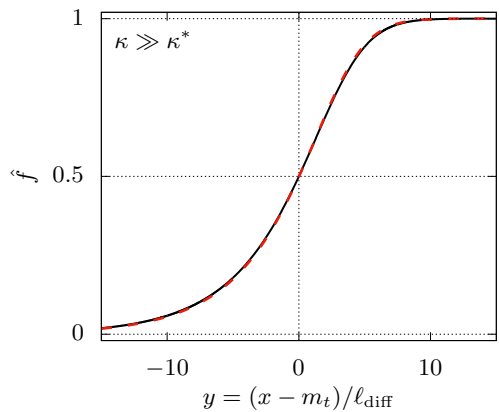}
    \caption{Radial profile of the late-time FKPP front at $\kappa \gg 1$, solution of Eq.~(\ref{eq:FKPP_diff}) traveling at the butterfly velocity $v$ given in Eq.~(\ref{eq:v_diff}). $m_t\sim vt$ is the location of the front.
    The dashed line is the numerical result obtained by solving Eqs.~(\ref{eq:coupled_PDEs_adim}) with $\kappa = 60$ and $\gamma=1$ up to time $\tau = 30$. The front thickness is of the order of $\ell_{\rm diff}:=v_\rmF \sqrt{\tau_{\rm sc}\tau_{\rm el}/2d} \ll \ell_{\rm sc}$. }    \label{fig:diffusive}
\end{figure}
We now delve into the overdamped regime $\kappa \gg 1$ where the inertial term of Eq.~(\ref{eq:FKPP}) may be neglected. Keeping only the leading-order terms the right-hand side
(see the details in~\autoref{app:diffu}), the scrambling dynamics are now governed by, back in terms of $\phi$ and in the original units,
\begin{align} \label{eq:FKPP_diff}
    \partial_t \phi - D_{\rm el} \nabx^2 \phi =  \phi(\phi^2-1) / \tau_{\rm sc},
\end{align}
where $1/\tau_{\rm sc}$ is the inelastic scattering rate and $D_{\rm el} := v_\rmF^2 \tau_{\rm el} /d$ is the elastic diffusion coefficient. The dependence on both $\boldsymbol{\phi}_1$ and $\gamma$ has dropped.
This FKPP equation is the non-integrable Newell-Whitehead equation which was first studied in the context of non-linear fluid mechanics~\cite{Newell1969}. Similar equations with diffusive terms were recently put forward to describe the dynamics of scrambling of quantum information~\cite{Aleiner2016,Swingle2017_ON,Scopelliti2019,Swingle2019,SwingleZhou2023}.
The traveling-wave solutions of Eq.~(\ref{eq:FKPP_diff}) propagate at a butterfly velocity, in the original units,
\begin{align} \label{eq:v_diff}
v = 2\sqrt{\frac{2}{d}} \sqrt{\frac{\tau_{\rm el}}{\tau_{\rm sc}}} v_\rmF \ll v_{\rmF}. 
\end{align}
Incidentally, this yields the relation 
\begin{align}
v^2/\lambda_{\rm FKPP} = 2 D_{\rm el}
\end{align}
which concretely connects scrambling dynamics on the one side to a measurable transport quantity on the other side~\cite{Lucas2017, Sachdev2017, Stanford2017, Moessner2018b}.
The front profile is conveniently computed by now measuring space in units of $l_\text{diff} := v_\rmF \sqrt{\tau_{\rm sc}\tau_{\rm el}/2d}\ll \ell_{\rm sc}$. The rescaled front profile $\hat f(y):=f(y \, l_\text{diff})$ is the solution to $2 \hat f'' +  4\hat f' = \hat f(1 - \hat f^2)$ with $\hat f(-\infty) = 0$, $\hat f(\infty)=1$, and we can require $\hat f(0) = 1/2$. It is a smooth monotonous function that we represent in Fig.~\ref{fig:diffusive}. The agreement with the numerical solutions of the PDEs~(\ref{eq:coupled_PDEs_adim}) computed in the diffusive regime ($\kappa = 60$) is excellent.

\paragraph*{Discussion.}
In the Fermi shock-wave regime, $\kappa  < \kappa^*$, it is still to be clarified whether the scrambling front discontinuity could be smoothened by corrections to the gradient approximation. 
In the FKPP regime, $\kappa  > \kappa^*$, we found that disorder can considerably reduce the butterfly velocity, corroborating the results of Refs.~\cite{Swingle2017_disorder,SwingleSachdev2017}: a realistic ratio at room temperatures $\kappa \approx 10^4$ yields $v \sim 10^{-2} v_\rmF \sim 10^4$~m/s, on par with the typical phonon-mediated sound velocity in metals.
Contrary to the shock-wave velocity, it cannot strictly be interpreted a (slower) effective speed of light since the small tail ahead of the front, while providing a quantum-chaotic exponential regime controlled by the inelastic scattering rate,
surreptitiously undermines the causality of the light-cone structure.
In both the Fermi shock-wave and the FKPP regimes, we found a sharp butterfly front. This is similar to what has been reported for models with a large local Hibert space such as the $O(N)$ or the Sachdev-Ye-Kitaev (SYK) models~\cite{Swingle2017_ON, Stanford2017}, but different from the diffusively broadening fronts obtained for random quantum circuits~\cite{Nahum2018,Piroli2020}. The precise conditions under which strong quantum fluctuations and strong disorder could generate relevant perturbations to the FKPP dynamics are still to be elucidated.

\paragraph*{Acknowledgments.}
%\begin{acknowledgments}
We would like to thank Alex Kamenev for his initial inspiration which was crucial in the development of this research project.
We are grateful to Giulio Biroli and Aavishkar A. Patel for insightful discussions.
C.A. acknowledges the support from the French ANR ``MoMA'' Project No. ANR-19-CE30-0020 and the support from the Project No. 6004-1 of the Indo-French Centre for the Promotion of Advanced Research (IFCPAR). 
This work received support from the U.S. National Science Foundation Grant No. NSF-DMR 2018358 (A.M.). 
%\end{acknowledgments}

\bibliography{main.bib}
\clearpage
\appendix

\onecolumngrid

\section{Kinetic equations of scrambling in a disordered interacting metal} \label{app:model}
\begin{figure}
    \centering
    \includegraphics[scale=3]{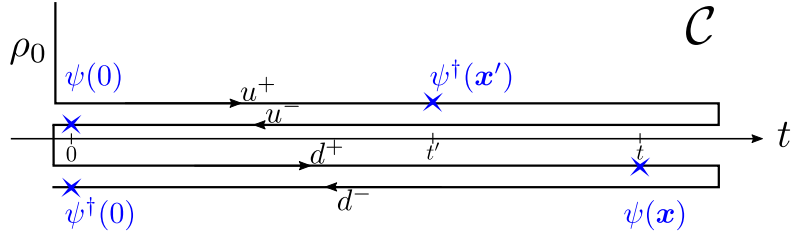}
    \caption{Two-world Keldysh contour $\mathcal{C}$: the theory is replicated into an ``up'' world dynamics, marked by the index $u$, and the ``down'' world dynamics marked by $d$. In blue: the insertion points of the OTOC $\mathcal{A}(t,\bx;t',\bx')  := \langle \mathrm{T}_\mathcal{C} \,    \psi_d^{-\dagger}(0,0)  \psi_{d}^{+}(t,\bx) \psi_u^{-}(0,0) \psi_{u}^{+\dagger}(t',\bx')\rangle$.
    }
    \label{fig:manyworld}
\end{figure}

In this Appendix, we derive the set of kinetic equations~(4) that govern the dynamics of quantum information scrambling in the disordered interacting metal given by the Hamiltonian in Eq.~(1).
More information on the procedure for the clean case can be found in Ref.~\cite{ABM}.

\subsection{Many-world formalism} \label{app:formalism}
We  work within the many-world formalism, sometimes referred to as the augmented Keldysh formalism. We refer the reader to Ref.~\cite{Aleiner2016} for a pedagogical introduction, see also Ref.~\cite{kamenev2023}.
Two-point correlators are defined as
\begin{align} \label{eq:GF}
    \rmi G_{\alpha\beta}^{ab}(t,\bx;t',\bx') &:= \langle \mathrm{T}_\mathcal{C} \psi_{\alpha}^a(t,\bx)  \psi_{\beta}^{b\dagger}(t',\bx') \rangle, 
\end{align}
with the Keldysh indices $a,b = +,-$, and the world indices $\alpha,\beta=u,d$. We set $\hbar = 1$.
$\mathrm{T}_\mathcal{C}$ is the time-ordering operator on the two-world contour $\mathcal{C}$ represented in Fig.~\ref{fig:manyworld}. The fermionic operators $\psi$ and $\psi^\dagger$ are written in the Heisenberg picture, and $\langle \ldots \rangle := \mathrm{Tr} \left[ \ldots \rho_0 \right]$ where $\rho_0$ is the initial density matrix with $\mathrm{Tr}\, \rho_0 = 1$.
The system is initially prepared in equilibrium at the temperature $T$ with the Gibbs state $\rho_0 \sim \rme^{-H/T}$ where we set $k_{\rm B} = 1$.
The equilibrium dynamics ensure that, in the absence of an external perturbation explicitly breaking this invariance, the intraworld dynamics is translational invariant in space and time, and one may work in Fourier space.
This is not the case for interworld quantities ($\alpha\neq\beta$) since we are precisely probing their response with respect to local perturbations. Therefore, we use the Wigner representation for the interworld components, \textit{e.g.}
\begin{align}
    G^{ab}_{\alpha \beta}(\omega,\bk;t,\bx) := \int\!\! \rmd \bx' \!\! \int \!\!\rmd t' \,  \rme^{\rmi (\omega t' - \bk \cdot \bx')} \,  G^{ab}_{\alpha\beta}\left( t+\frac{t'}{2} ,\bx+\frac{\bx'}{2}; t-\frac{t'}{2},\bx-\frac{\bx'}{2}\right).
\end{align}
To alleviate the notations, we drop the intraworld indices, \textit{e.g.} $G_{uu} = G_{dd} = G$, and we reserve the double indices $\alpha\beta$ to interworld quantities, \textit{i.e.} by $\alpha\beta$ we mean $\alpha\neq\beta$ specifically.
We work in the Keldysh basis with the retarded ($R$) and Keldysh ($K$) Green's functions which, after using the properties of the contour $\mathcal{C}$, read
\begin{align}
\left\{
\begin{array}{rl}
    G^R &= G^{++} - G^{+-} \\
    G^K &=  G^{++} + G^{--}\
    \end{array}
\right.
\mbox{ and }
\left\{
\begin{array}{rl}
    G^R_{\alpha\beta} &= 0 \\
    G^K_{\alpha\beta} &= 2 G_{\alpha\beta}^{-+}.
    \end{array}
\right.
\end{align}
The intraworld Schwinger-Dyson equations yield the standard (single-world) theory
\begin{align}
\left\{ \label{eq:Dyson_aa}
\begin{array}{rl}
         G^R(\omega, \bk) &= \left[ \omega - \epsilon(\bk) - \Sigma^R(\omega,\bk) \right]^{-1}\\
    G^K(\omega, \bk) & =  |G^R(\omega,\bk)|^2 \Sigma^K(\omega,\bk).
\end{array}
\right.
\end{align}
The interworld Schwinger-Dyson equations read
\begin{align}
\left\{ \label{eq:Dyson_ab}
\begin{array}{rl}
G_{\alpha\beta}^R(\omega,\bk;t,\bx) &= 0 \\
    G^K_{\alpha\beta}(\omega,\bk;t,\bx) & =  G^R(\omega,\bk;t,\bx) \star \Sigma^K_{\alpha\beta}(\omega,\bk;t,\bx)\star G^R(\omega,\bk;t,\bx)^*,
    \end{array}
\right.
\end{align}
where we introduced the Moyal product $\star :=\exp\left[\frac{\rmi}{2} (
\overleftarrow{\partial_\omega} \overrightarrow{\partial_t} 
\!-\! \overleftarrow{\nabk} \cdot \overrightarrow{\nabx}
\!-\! \overleftarrow{\partial_t} \overrightarrow{\partial_\omega}
\!+\!\overleftarrow{\nabx} \cdot \overrightarrow{\nabk}
)  \right] $ where the left (right) arrow designates a derivative operator acting on the left (right) of the star symbol.
The self-energies (the $\Sigma$'s) entering the Schwinger-Dyson equations (\ref{eq:Dyson_aa}) and (\ref{eq:Dyson_ab}) will be discussed below in App.~\ref{app:self}.

\paragraph*{Connection to scrambling.\\}
Let us briefly outline the connection between the interworld two-point Green's functions and the four-point OTOCs which are commonly used to probe the dynamical signatures of quantum chaos. An example of such an OTOC is
\begin{align} \label{eq:OTOC}
\mathcal{A}(t,\bx;t',\bx') :=& \mathrm{Tr}\, \left[ \psi^\dagger(0) \ \rme^{\rmi H t} \psi(\bx) \rme^{-\rmi H t} \ \psi(0) \ \rme^{\rmi H t'} \psi^\dagger(\bx') \ \rme^{-\rmi H t'} \rho_0 \right] \\
=& 
    \langle \mathrm{T}_\mathcal{C} \,    \psi_d^{-\dagger}(0,0)  \psi_{d}^{+}(t,\bx) \psi_u^{-}(0,0) \psi_{u}^{+\dagger}(t',\bx')\rangle.
\end{align}
At late times, we expect the decoupling
\begin{align}
    \mathcal{A}(t,\bx;t',\bx') \simeq 
     n_0 \, \rmi G_{du}^{++}(t,\bx;t',\bx') \propto G^K_{du}(t,\bx;t',\bx'),
\end{align}
with $n_0 := \mathrm{Tr}\, \left[ \psi^\dagger(0) \psi(0)  \rho_0 \right]$ and where the interworld Green's function is computed as the solution of the interworld Schwinger-Dyson equations in the presence of a local perturbation to the correlated world solution at $\bx = 0$ and time $t=0$.

\subsection{Self-energy contributions} \label{app:self}
There are two contributions to the electronic self-energy: elastic scattering on the disordered potential and inelastic scattering on the Cooperons. We refer the reader to Chap. 8 in Ref.~\cite{rammer_2007}, specifically around Eqs~(8.180) and~(8.181), for a detailed discussion of the self-energy contributions to the (intraworld) kinetic equations in a very similar setting. We follow the same approach for the interworld theory and
\begin{align}
\left\{
\begin{array}{rl}
    \Sigma^{R} &= \Sigma^{R}_{\rm dis}  + \Sigma^{R}_{U}    \\[\smallskipamount]
    \Sigma^{K} &= \Sigma^{K}_{\rm dis}  + \Sigma^{K}_{U}   
    \end{array}
\right.
\mbox{ and }
    \left\{
\begin{array}{rl}
    \Sigma^{R}_{\alpha\beta} &= \Sigma^{R}  \\[\smallskipamount]
    \Sigma^{K}_{\alpha\beta} &= \Sigma^{K}_{{\rm dis}\, \alpha\beta}  + \Sigma^{K}_{U\, \alpha\beta} ,
    \end{array}
\right.
\end{align}
where the disorder contribution $\Sigma_{\rm dis}$ and the interacting contribution $\Sigma_{U}$ are detailed below. It is to be noted that both contributions to the self-energy feed each other. In particular, the interacting contribution depends implicitly on the disorder strength.

\bigskip

\paragraph*{Disorder contribution.\\}
Expanding to the second order in the coupling to the disordered potential, we obtain the self-energy contribution
\begin{align}
\left\{
\begin{array}{rl}
    \Sigma^{R/K}_{{\rm dis}}(\omega,\bk) &= g \sum_{\bk'}  G^{R/K}(\omega,\bk')  \\[\smallskipamount]
    \Sigma^{K}_{{\rm dis} \, \alpha\beta}(\omega,\bk;t,\bx) &= g \sum_{\bk'} 
    G^{K}_{\alpha\beta}(\omega,\bk';t,\bx),
\end{array}
\right.
\end{align}
where $g \delta(\boldsymbol{x}-\bx') := \langle V(\bx) V(\bx') \rangle - \langle V(\bx) \rangle^2$.

\bigskip

\paragraph*{Interaction contribution.\\}
We treat the finite electronic interaction in the Cooper channel by means of the random phase approximation (RPA). The intraworld self-energy contributions ($\alpha=\beta$) read
\begin{align}
\left\{  \label{eq:Sigma}
\begin{array}{rl} 
\,\mathrm{Im}\,\Sigma_{U}^R(\omega,\bk) \!\!\!\!&= \displaystyle - \frac{1}{2} \sum_{\bk'} \int \frac{\rmd \omega'}{2\pi} \bigg{\{}
\,\mathrm{Im}\, D^R(\omega',\bk') \rmi  G^K(\omega'-\omega,\bk'-\bk) 
-  \rmi D^K(\omega',\bk')\,\mathrm{Im}\, G^R(\omega'-\omega,\bk'-\bk) \bigg{\}}  \\[\medskipamount]
\Sigma_{\rm U}^K(\omega,\bk) \!\!\!\!&=  2\rmi \, F(\omega) \,\mathrm{Im}\,\Sigma^R(\omega,\bk) ,
\end{array}
\right.
\end{align}
where $F(\omega) := \tanh(\omega/2T)$ is dictated by the (fermionic) fluctuation-dissipation theorem.
Within the RPA in the particle-particle channel, the Cooperon propagators are given by \begin{align}
\left\{  \label{eq:D_R}
\begin{array}{l}  
D^R(\omega,\bk) =   \left[ U^{-1} -  \Pi^R(\omega,\bk) \right]^{-1} \\[\medskipamount]
D^K(\omega,\bk)  = 2\rmi \, P(\omega) \,\mathrm{Im}\, D^R(\omega,\bk) ,
\end{array}
\right.
\end{align}
where $P(\omega) := \coth(\omega/2T)$ is dictated by the (bosonic) fluctuation-dissipation theorem and with the polarization bubbles read
\begin{align}
\left\{  \label{eq:Pi}
\begin{array}{rl}  
\,\mathrm{Im}\, \Pi^R(\omega,\bk) \!\!\!\!&=  \displaystyle \sum_{\bk'} \int \frac{\rmd \omega'}{2\pi}  \rmi G^K(\omega',\bk')  \,\mathrm{Im}\, G^R(\omega-\omega',\bk-\bk') \\[\medskipamount]
\Pi^K(\omega,\bk) \!\!\!\!&= 2\rmi \, P(\omega) \,\mathrm{Im}\,\Pi^R(\omega,\bk) ,
\end{array}
\right.
\end{align}
Close to the superconducting transition, the near-critical Cooperon propagator reads
\begin{align}
    D^R(\omega,\bk) \approx \frac{-1/\rho_\rmF}{r - \rmi a {\omega}/{T}+ \xi^2 k^2 + \ldots}, 
\end{align}
where $\rho_\rmF$ is the density of states at the Fermi energy, and the parameter $r \propto (T - T_{\rm c})/T_{\rm c}$ is the detuning from the critical temperature $T_{\rm c}$. In addition, the remaining parameters are all positive with $a \sim \mathcal{O}(1)$, $\xi^2 \sim  v_\rmF^2 /T^2$, where $v_\rmF$ is the Fermi velocity.
At criticality $r\to0$, the Cooperon becomes soft with diverging length scale $l\sim 1/r^\nu$ and timescale $\sim l^z$ (here $\nu = 1/2$, $z=2$), and the propagator is singular at $\omega = k =0$.

\bigskip

The interworld components read
\begin{align}
\Sigma_{U\, \alpha\beta}^K(\omega,\bk;t,\bx) &=  -\frac{\rmi}{2} \sum_{\bk'}  \int \frac{\rmd \omega'}{2\pi} D_{\alpha\beta}^K(\omega',\bk';t,\bx)  G_{\beta\alpha}^K(\omega'-\omega,\bk'-\bk;t,\bx),
\end{align}
with
\begin{align}
D_{\alpha\beta}^K(\omega,\bk;t,\bx) &=  D^R(\omega,\bk) \star \Pi_{\alpha\beta}^K(\omega,\bk;t,\bx) \star  D^R(\omega,\bk)^*,\\
\Pi_{\alpha\beta}^K(\omega,\bk;t,\bx) &=  \frac{\rmi}{2} \sum_{\bk'}  \int \frac{\rmd \omega'}{2\pi} G_{\alpha\beta}^K(\omega',\bk';t,\bx) G_{\alpha\beta}^K(\omega-\omega',\bk-\bk';t,\bx).
\end{align}

\bigskip

\subsection{Interworld kinetics}
We start from the quantum kinetic equation which is a rewrite of the Schwinger-Dyson equations~(\ref{eq:Dyson_ab}). See, \textit{e.g.}, a standard derivation in Ref~\cite{kamenev2023}. The interworld Keldysh Green's function ($\alpha\neq\beta$) is parametrized in terms of the interworld distribution function $F_{\alpha\beta}$:
\begin{align}
    G^K_{\alpha \beta}(\omega,\bk;t,\bx)
    =  G^R(\omega,\bk)  \star F_{\alpha \beta}(\omega,\bk;t,\bx) - F_{\alpha \beta}(\omega,\bk;t,\bx)  \star G^A(\omega,\bk),
    \label{eq:F_param}
\end{align}
To alleviate the notations, we now drop the explicit dependence of $\bx$ and $t$ except when this obscures the meaning.

\paragraph*{Gradient approximation.} Expanding the Moyal products to first non-trivial orders, the interworld quantum kinetic equation reads
\begin{align} \label{eq:inter_kinetic}
      \left[ \partial_t + \boldsymbol{v}_\bk \cdot \nabx \right] F_{\alpha\beta}(\omega,\bk ) &= I_{\alpha \beta}(\omega,\bk ),
\end{align}      
with the collision integral
\begin{align}
     I_{\alpha \beta}(\omega,\bk ) & =  2 \,  {\rm Im} \Sigma^{R}(\omega,\bk) \,F_{\alpha \beta}(\omega,\bk) + \rmi\Sigma^K_{\alpha\beta}(\omega,\bk).
     \end{align}
Using the expressions for the self-energies in App.~\ref{app:self}, we have
     \begin{align} 
         I_{\alpha \beta}(\omega,\bk )  & = 2 \sum_{\bk'} g(\bk-\bk') \mathrm{Im}\, G^{R}(\omega,\bk')  \left[ F_{\alpha \beta}(\omega,\bk) - F_{\alpha \beta}(\omega,\bk') \right]   \nonumber \\
       & \quad +  2
  \sum_{\bk'\bk''} \int \frac{\rmd \omega'}{2\pi} \frac{\rmd \omega''}{2\pi} 
  |D^R(\omega',\bk')|^2 
  \nonumber \\
  & \quad \qquad \times   
  \,\mathrm{Im}\, G^R(\omega'',\bk'') \,\mathrm{Im}\, G^R(\omega'-\omega'',\bk'-\bk'') \,\mathrm{Im}\, G^R(\omega'-\omega,\bk'-\bk) \nonumber \\
  & \quad \qquad \times \bigg{\{}  \left[ \tanh\left(\frac{\omega''}{2T}\right) +\tanh\left(\frac{\omega'-\omega''}{2T}\right) \right] \left[ \coth\left(\frac{\omega'}{2T}\right) +\tanh\left(\frac{\omega-\omega'}{2T}\right) \right]  F_{\alpha\beta}(\omega,\bk) \nonumber \\ 
  &\quad \qquad\qquad  + 
   F_{\alpha\beta}(\omega'',\bk'')
   F_{\alpha\beta}(\omega'-\omega'',\bk'-\bk'')
   F_{\beta\alpha}(\omega'-\omega,\bk'-\bk) \bigg{\}} . \label{eq:Coll_withoutP} 
\end{align}
One may check that the collision integral vanishes identically at the uncorrelated world solution
\begin{align}
    F^{\rm uncorr}_{\alpha\beta}(\omega,\bk) = 0,
\end{align}
as well as at the correlated world solution 
\begin{align}
 F^{\rm corr}_{\alpha\beta}(\omega,\bk) = \epsilon_{\alpha\beta} + \tanh\left( \omega/2T \right),
\end{align}
with $\epsilon_{du} = 1$ and $\epsilon_{ud} = -1$.

\bigskip
\paragraph*{Quasi-particle approximation.} In all practical instances, $F_{\alpha\beta}(\omega,\bk)$ appears multiplied by the density of states $\,\mathrm{Im}\, G^R(\omega,\bk)$, see \textit{e.g.} Eq.~(\ref{eq:Coll_withoutP}). When quasi-particles are well defined, with a dispersion relation $\epsilon_\bk$, the density of states is sharply peaked around $\omega = \epsilon_\bk$ and one may seamlessly exchange $F_{\alpha\beta}(\omega,\bk)$ with the on-shell quasi-particle distribution function $\tilde F_{\alpha\beta}(\bk) := F_{\alpha\beta}(\omega=\epsilon_\bk,\bk)$. From now on, we use the tilde notation to denote the on-shell prescription $\omega = \epsilon_\bk$.

\bigskip
\paragraph*{Partial-wave ansatz.}
The kinetic equation in~(\ref{eq:inter_kinetic}), with its collision integral in Eq.~(\ref{eq:Coll_withoutP}), is considerably simplified by working with the following ansatz which implements, altogether, the truncation to the first terms of a partial-wave expansion in the momentum space, the quasi-particle approximation, and the focusing on the physics at the Fermi surface. Restoring the explicit dependence on $\bx$ and $t$, it reads
\begin{align} \label{eq:Ansatz}
    \tilde F^{\rm ansatz}_{\alpha\beta}(t,\bx;\bu_\bk)  &= \left[ \phi(t,\bx)  + \bu_\bk \cdot \boldsymbol{\phi}_1(t,\bx)  \right]  \, \tilde F_{\alpha\beta}^{\rm corr}(\bk)\lvert_{k\to k_\rmF} \\
    & = \epsilon_{\alpha\beta} \left[ \phi(t,\bx)  + \bu_\bk \cdot \boldsymbol{\phi}_1(t,\bx)  \right],
\end{align}
with the unit vector $\bu_\bk := \bk/k$ and where $\phi$ and $\boldsymbol{\phi}_1$ are real scalar and vector fields coding, respectively, for the isotropic component and the first anisotropic corrections to the interworld distribution function $\tilde F_{\alpha\beta}(\bk) \big{\lvert}_{k \to k_\rmF}$.
Conversely, this amounts to performing a projection on the first two terms of a partial-wave expansion:
\begin{align} \label{eq:phi_phi1}
 \left\{ \begin{array}{rl}
\phi(t,\bx) \!\!\!\! &= \displaystyle \frac{1}{S_{d-1}} \epsilon_{\alpha\beta} \int\rmd\Omega_\bk F_{\alpha\beta}(t,\bx;\epsilon_\bk,\bk) \Big{\lvert}_{k \to k_\rmF} \\[\medskipamount]
     \boldsymbol{\phi}_1(t,\bx) \!\!\!\! &= \displaystyle \frac{d}{S_{d-1}} \epsilon_{\alpha\beta} \int\rmd\Omega_\bk \bu_\bk F_{\alpha\beta}(t,\bx;\epsilon_\bk,\bk)  \Big{\lvert}_{k \to k_\rmF},
    \end{array}
    \right.
\end{align}
where $\rmd \Omega_\bk$ is the elementary solid angle in the direction of $\bk$, and $S_{d-1} := \int \rmd \Omega_\bk$ is the surface area of the $d-1$-sphere with unit radius. One has $\int \rmd \Omega_\bk u_\bk^i u_\bk^j =$ $\delta_{ij} \, S_{d-1}/d$.
From Eq.~(\ref{eq:inter_kinetic}), we obtain
\begin{align} \label{eq:pre_proj}
     &\left[ \partial_t \phi + \bu_{\bk} \cdot \partial_t \boldsymbol{\phi}_1 + v_\rmF \left(  \bu_\bk \cdot \nabx \right) \left( \phi   +   \bu_\bk \cdot \boldsymbol{\phi}_1 \right) \right] \epsilon_{\alpha\beta} =    \tilde I_{{\rm dis}\, \alpha\beta}(\bk)\lvert_{k\to k_\rmF} + \tilde I_{U\, \alpha\beta}(\bk)\lvert_{k\to k_\rmF},
\end{align}
with
\begin{align}
    \tilde I_{{\rm dis}\, \alpha\beta}(\bk) & = g \boldsymbol{\phi}_1 \cdot \sum_{\bk'}  \mathrm{Im}\, G^R(\epsilon_\bk,\bk') \left[\bu_\bk \tilde F_{\alpha\beta}^{\rm corr}(\bk) -  \bu_{\bk'} \tilde F_{\alpha\beta}^{\rm corr}(\bk') \right].
\end{align}
At the Fermi surface, $k\to k_\rmF$, the term  $\mathrm{Im}\, G^R(\epsilon_\bk,\bk')$ selects momenta $\bk'$ which also live at the Fermi surface.
We obtain
\begin{align}
    \tilde I_{{\rm dis}\, \alpha\beta}(\bk) \lvert_{k\to k_\rmF} & = - \epsilon_{\alpha\beta} \pi g \rho_\rmF  \, \boldsymbol{\phi}_1 \cdot \bu_\bk \\
    &= - \epsilon_{\alpha\beta}  \boldsymbol{\phi}_1 \cdot \bu_\bk / \tau_{\rm el},
\end{align}
where we made use of $\tilde F_{\alpha\beta}(\bk)\lvert_{k\to k_\rmF} = \epsilon_{\alpha\beta}$ with $\epsilon_{du} = -\epsilon_{ud} = 1$, and $\rho_\rmF$ is the total density of states at the Fermi energy.
In the last line, we introduced the elastic scattering rate
\begin{align}
    \frac{1}{\tau_{\rm el}} :=  \pi g \rho_\rmF.
\end{align}
The use of the Born approximation is legitimate far from the localization regimes, $1/{\tau_{\rm el}}\ll  E_{\rmF}$.
On the interaction side, using the results of Ref.~\cite{ABM}, we have
\begin{align}
    \tilde I_{U\, \alpha\beta}(\bk)  \lvert_{k\to k_\rmF}=  \epsilon_{\alpha\beta}  \phi (\phi^2-1)  / \tau_{\rm sc} 
    + \epsilon_{\alpha\beta}  (\boldsymbol{\phi}_1 \cdot \bu_\bk)  (\gamma\phi^2-1)  / \tau_{\rm sc} + \ldots
\end{align}
where the $\ldots$ stand for higher-order spherical harmonics and we introduced the (temperature and disorder dependent) inelastic scattering rate 
\begin{align}
\frac{1}{\tau_{\rm sc}} := -2 \,\mathrm{Im}\, \tilde \Sigma^R(k_\rmF) , \label{eq:def_tds}
 \end{align}
 with, explicitly, 
   \begin{align}
       {\rm Im}\, \tilde \Sigma_U^{R}(\bk) 
     & = 
  \sum_{\bk'\bk''} \int \frac{\rmd \omega'}{2\pi} \frac{\rmd \omega''}{2\pi} 
  |D^R(\omega',\bk')|^2 
  \,\mathrm{Im}\, G^R(\omega'',\bk'') \,\mathrm{Im}\, G^R(\omega'-\omega'',\bk'-\bk'') \,\mathrm{Im}\, G^R(\omega'-\epsilon_\bk,\bk'-\bk) \nonumber \\
  & \qquad \qquad \qquad \qquad \times  \left[ \tanh\left(\frac{\omega''}{2T}\right) +\tanh\left(\frac{\omega'-\omega''}{2T}\right) \right] \left[ \coth\left(\frac{\omega'}{2T}\right) +\tanh\left(\frac{\epsilon_\bk-\omega'}{2T}\right) \right], 
  \end{align}
and where the parameter $0 < \gamma \leq 1$ quantifies the distance to criticality ($\gamma = 1$ at criticality).

We now project Eq.~(\ref{eq:pre_proj}) onto the first two terms of the partial-wave expansion, using
\begin{align}
 \left\{ 
 \begin{array}{rl}
  \int \rmd \Omega_\bk \tilde I_{{\rm dis}\, \alpha\beta}(\bk)  \lvert_{k\to k_\rmF} & = 0
 \\[\medskipamount]
     \frac{d}{S_{d-1}}   \int \rmd \Omega_\bk \bu_\bk \tilde I_{{\rm dis}\, \alpha\beta}(\bk)  \lvert_{k\to k_\rmF} & = - \epsilon_{\alpha\beta} \boldsymbol{\phi}_1 / \tau_{\rm el}
\end{array}
     \right.
\mbox{ and }
 \left\{ 
 \begin{array}{rl}
  \int \rmd \Omega_\bk \tilde I_{U\, \alpha\beta}(\bk)  \lvert_{k\to k_\rmF} & = \epsilon_{\alpha\beta} \phi(\phi^2-1)/\tau_{\rm sc}
\\[\medskipamount]
     \frac{d}{S_{d-1}}   \int \rmd \Omega_\bk \bu_\bk \tilde I_{U\, \alpha\beta}(\bk)  \lvert_{k\to k_\rmF} & =  \epsilon_{\alpha\beta} \boldsymbol{\phi}_1 (\gamma \phi^2-1) / \tau_{\rm sc}.
     \end{array}
     \right.
\end{align}
Finally, we obtain the set of coupled PDEs~(4) which govern the kinetics of quantum information scrambling,
\begin{align}
\left\{  
\begin{array}{rl}  
     \partial_t \phi +  \frac{v_\rmF}{d} \nabx   \cdot \boldsymbol{\phi}_1 \!\!\!\!& =  \phi (\phi^2-1)/\tau_{\rm sc}
 \\[\medskipamount]
          \partial_t  \boldsymbol{\phi}_1 +  v_\rmF \nabx \phi  \!\!\!\!& = \boldsymbol{\phi}_1  (\gamma\phi^2-1) /\tau_{\rm sc} - \boldsymbol{\phi}_1 / \tau_{\rm el}.
\end{array}
\right.
\end{align}

\section{Light-cone structure of scrambling dynamics} \label{app:ballistic}

Here below, we carefully demonstrate that the solutions of the coupled PDEs~(5) in $d=1$ are strictly causal in the sense that they are bounded by a light cone which propagates at velocity $v_{\rm lc} = 1$ (\textit{i.e.}\@ $v_{\rm lc} = v_\rmF / \sqrt{d}$ in the original units). In other words, for any $X$ and any $\tau>0$, the values of $\phi(\tau,X)$ and $\phi_1(\tau,X)$ depend only on the initial conditions $\phi_0(z)$ and ${\phi_{1}}_0(z)$ in the range $z\in[X-\tau,X+\tau]$. Then, with initial conditions such that $\phi_0(X)=1$ and ${\phi_{1}}_0(X)=0$ for all $|X|> R_0$, as studied in the paper, the solution strictly satisfies
\begin{equation}
\phi(\tau,X)=1\quad\text{and}\quad\phi_1(\tau,X)=0\quad\text{for $|X|>R_0+\tau$}.
\end{equation}
Let us first recall the system of equations~(5) in dimension $d=1$:
\begin{equation}
\label{systemqq1}
\left\{
\begin{array}{rl}
\partial_\tau \phi   + \partial_X \phi_1 &= \phi(\phi^2-1)\\
\partial_\tau \phi_1 + \partial_X \phi &=\phi_1\left(\gamma \phi^2-1-\kappa\right).
\end{array}
\right.
\end{equation}
where $0\le\gamma\le1$  and $\kappa\ge0$. 
More generally, the light cone property can be proved to hold for any system of the form 
$\partial_\tau \phi   + \partial_X \phi_1=f(\phi,\phi_1)$,
$\partial_\tau \phi_1 + \partial_X \phi = g(\phi,\phi_1)$ with 
``reasonable'' (say, polynomial) functions $f$ and $g$, and for any bounded
initial condition. However, in order to simplify the discussion, we only
consider the specific case~\eqref{systemqq1}
and  initial conditions such that
\begin{equation}
0\le\phi_0(X)\le1,\qquad{\phi_{1}}_0(X)=0.
\label{eq:initcondagain}
\end{equation}
For such an initial condition, we also show that the solution satisfies the following properties:
\begin{itemize}
\item $0\le \phi(\tau,X)\le 1$ for all $X$ and $\tau\ge0$,
\item if $\phi_0(X)$ and $\tilde\phi_0(X)$ are two initial conditions satisfying~\Eq{eq:initcondagain} such that $\phi_0(X)\le \tilde\phi_0(X)$ for all $X$, then the corresponding solutions satisfy $\phi(\tau,X)\le \tilde\phi(\tau,X)$ for all $\tau>0$ and all $X$.
\end{itemize}

\subsection{A construction of the solutions \texorpdfstring{$\phi$}{φ}
 and \texorpdfstring{$\phi_1$}{φ₁}}

To show these properties, we rewrite~\Eq{systemqq1} in terms of $\varphi:= \phi+  \phi_1$ and $\psi:=\phi-\phi_1$:
\begin{equation}
\begin{cases}
\partial_\tau \varphi   + \partial_X \varphi 
= \varphi (\gamma \phi^2-1-\kappa )+(1-\gamma)\phi^3+\kappa\phi \\
\partial_\tau \psi - \partial_X \psi =\psi(\gamma \phi^2-1-\kappa )+(1- \gamma)\phi^3+\kappa\phi.
\end{cases}
\label{systemSD}
\end{equation}
Going respectively into the $X=z+\tau$ and the $X=z-\tau$ frame, this
is equivalent to
\begin{equation}
\begin{cases}
\dfrac{\rmd}{\rmd \tau} \varphi(\tau,z+\tau)
= \Big[\varphi (\gamma \phi^2-1-\kappa )+(1-\gamma)\phi^3+\kappa\phi\Big]_{X=z+\tau} \\[2ex]
\dfrac{\rmd}{\rmd \tau} \psi(\tau,z-\tau)
= \Big[\psi (\gamma \phi^2-1-\kappa )+(1-\gamma)\phi^3+\kappa\phi\Big]_{X=z-\tau}.
\end{cases}
\end{equation}
Then, this is also equivalent to the integral equations
\begin{equation*}
\begin{cases}
\varphi(\tau,z+\tau)
= \displaystyle\rme^{\int_0^\tau\rmd s\,[\gamma \phi^2(s,z+s)-1-\kappa ]}\varphi_0(z)+\int_0^\tau\rmd u\,\rme^{\int_u^\tau\rmd s\,[\gamma \phi^2(s,z+s)-1-\kappa ]}\big[(1-\gamma)\phi(u,z+u)^3+\kappa\phi(u,z+u)\big] \\[2ex]
\psi(\tau,z-\tau)
= \displaystyle\rme^{\int_0^\tau\rmd s\,[\gamma \phi^2(s,z-s)-1-\kappa ]}\psi_0(z)+\int_0^\tau\rmd u\,\rme^{\int_u^\tau\rmd s\,[\gamma \phi^2(s,z-s)-1-\kappa ]}\big[(1-\gamma)\phi(u,z-u)^3+\kappa\phi(u,z-u)\big],
\end{cases}
\end{equation*}
\vspace{-8.5ex}
\begin{equation}
\label{egint}
\end{equation}
where $\varphi_0(z)=\phi_0(z)+{\phi_{1}}_0(z)$ and $\psi_0(z)=\phi_0(z)-{\phi_{1}}_0(z)$. Note that, from~\Eq{eq:initcondagain}, $\varphi_0(z)=\psi_0(z)=\phi_0(z)$, but we only need to assume that $\varphi_0$ and $\psi_0$ are both in $[0,1]$, \textit{i.e.}\@ that $\phi_0\pm{\phi_{1}}_0\in[0,1]$.

Now, define recursively $\varphi_n(\tau,X)$ and $\psi_n(\tau,X)$ by
\begin{align}
\begin{cases}
\varphi_1(\tau,X):=0 \\ 
\psi_1(\tau,X):=0,
\end{cases}    
\end{align}
and
\begin{equation*}
\begin{cases}
\varphi_{n+1}(\tau,z+\tau):= \displaystyle\rme^{\int_0^\tau\rmd s\,[\gamma \phi_n^2(s,z+s)-1-\kappa ]}\varphi_0(z)+\int_0^\tau\rmd u\,\rme^{\int_u^\tau\rmd s\,[\gamma \phi_n^2(s,z+s)-1-\kappa ]}\big[(1-\gamma)\phi_n(u,z+u)^3+\kappa\phi_n(u,z+u)\big]\\[2ex]
\psi_{n+1}(\tau,z-\tau)
:= \displaystyle\rme^{\int_0^\tau\rmd s\,[\gamma \phi_n^2(s,z-s)-1-\kappa ]}\psi_0(z)+\int_0^\tau\rmd u\,\rme^{\int_u^\tau\rmd s\,[\gamma \phi_n^2(s,z-s)-1-\kappa ]}\big[(1-\gamma)\phi_n(u,z-u)^3+\kappa\phi_n(u,z-u)\big],
\end{cases}
\end{equation*}
\vspace{-8.5ex}
\begin{equation}
\label{recur}
\end{equation}
where $\phi_n(\tau,X):=\frac12\big[\varphi_n(\tau,X)+\psi_n(\tau,X)\big].$
We shall show in~\autoref{app:conv} that, for any choice of $T>0$, the series of functions $\varphi_n$ and $\psi_n$ converge uniformly over all $X$ and all $\tau\in[0,T]$  as $n\to\infty$. The limits $\varphi$ and $\psi$ of these convergent series must then satisfy~\Eq{egint} for all $X$ and all $\tau$, which means that they are the solution to the system~\eqref{systemSD}.

\subsection{Properties of the solution} \label{app:properties_sol}
Once the convergence of $\varphi_n$ and $\psi_n$ is proved, we have a construction of the solution $\{\varphi,\psi\}$ to the system~\eqref{systemSD}. This construction allows us to prove all the properties we claimed:
\begin{description}
\item[Light cone property]
We observe by recursion that $\varphi_n$ and $\psi_n$ satisfy the light cone property. Indeed, it is obvious for $n=1$; assume now that the property holds for $\varphi_n$ and $\psi_n$, then $\phi_n(u,z+u)$ (and, similarly, $\phi_n(s,z+s)$) in the right hand side of~\Eq{recur} only depends on the initial condition in the interval $[z,2u]\in[z,2\tau]$.  This implies that $\varphi_{n+1}(\tau,z+\tau)$ only depends on the initial condition in the interval $[z,2\tau]$, and this is the light cone property for $\varphi_{n+1}$ (and, similarly, for $\psi_{n+1}$).

Then, going to the limit $n\to\infty$, we obtain that $\varphi$ and $\psi$, and then $\phi$ and $\phi_1$, satisfy the light cone property.

\item[Comparison property] Let us consider two initial conditions $\{\varphi_0(z),\psi_0(z)\}$ and $\{\tilde \varphi_0(z),\tilde \psi_0(z)\}$, as well as $\{\varphi_n,\psi_n,\phi_n=\tfrac12(\varphi_n+\psi_n)\}$ and $\{\tilde \varphi_n,\tilde \psi_n,\tilde \phi_n=\tfrac12(\tilde \varphi_n+\tilde \psi_n)\}$, the functions respectively obtained from these initial conditions with the recursion~\eqref{recur}. If $\varphi_0\le\tilde \varphi_0$ and $\psi_0\le\tilde\psi_0$, it is clear by recursion from~\Eq{recur} that, for all $n$,  one has $\varphi_n\le\tilde \varphi_n$, $\psi_n\le\tilde\psi_n$, and then $\phi_n\le\tilde\phi_n$. Note that argument makes use of $0\le\gamma\le1$ and $\kappa\ge0$.

Going to the limit $n\to\infty$, we obtain that $\varphi\le\tilde \varphi$, $\psi\le\tilde \psi$, and then $\phi\le\tilde\phi$.

\item[Bounds on \boldmath $\phi$\,] If $\varphi_0\ge0$ and $\psi_0\ge0$, it is immediately clear that, by recursion, $\varphi_n\ge0$ and $\psi_n\ge0$, which of course also implies that $\phi_n\ge0$. With slightly more work, it is also clear that if $\varphi_0\le1$ and $\psi_0\le1$, then by recursion $\varphi_n\le1$ and $\psi_n\le1$ (and $\phi_n\le1$). Note that argument makes use of $0\le\gamma\le1$ and $\kappa\ge0$.

Taking the limit $n\to\infty$, we obtain that $0\le \varphi\le 1$ and $0\le \psi\le 1$ (and then $0\le\phi\le1$) if $0\le \varphi_0\le1$ and $0\le\psi_0\le 1$.

Another way to obtain the same result would be to use the comparison property: observing that $\{\varphi=0,\psi=0\}$ and $\{\varphi=1,\psi=1\}$ are two solutions of~\eqref{systemSD}, then any initial condition ``in between'' must lead to a solution remaining between 0 and 1.

\end{description}

\subsection{Convergence of  \texorpdfstring{$\varphi_n$}{φ\_n} and \texorpdfstring{$\psi_n$}{ψ\_n}}
\label{app:conv}
Let us now explain why the series $\varphi_n$ and $\psi_n$ converge as $n\to\infty$. To simplify the argument, let us assume that the initial condition satisfies $0\le\varphi_0\le1$ and $0\le\psi_0\le1$; this implies from~\autoref{app:properties_sol} that for all $n$, $0\le\varphi_n\le1$, $0\le\psi_n\le1$, and therefore $0\le\phi_n\le1$. 
Let us introduce
\begin{equation}
\Delta_n(\tau)=\sup_{X\in\mathbb R}
\big|\varphi_{n+1}(\tau,X)-\varphi_n(\tau,X)\big|
 +\sup_{X\in\mathbb R} \big|\psi_{n+1}(\tau,X)-\psi_n(\tau,X)\big|.
\end{equation}
Below, we shall show that $\Delta_n(\tau)\to0$ sufficiently quickly as $n\to\infty$ to ensure it is summable; specifically, we shall show that, for every $T>0$, one has
\begin{equation}
\sum_{n\ge1}  \sup_{\tau\in[0,T]}\Delta_n(\tau) <\infty.
\label{eq:goal}
\end{equation}
Then, writing
\begin{equation}
\varphi_n(\tau,X)=\sum_{p=1}^{n-1}[\varphi_{p+1}(\tau,X)-\varphi_p(\tau,X)],
\end{equation}
(remember that $\varphi_1(\tau,X)=0$), we see that the series is convergent, as the running term is bounded by $\Delta_p(\tau)$. This proves that $\varphi_n$ (and, similarly, $\psi_n$) converges uniformly over  all $X$ and all $\tau\in[0,T]$ to a limiting $\varphi$, as required.

To prove~\Eq{eq:goal}, we use~\Eq{recur} to bound $\varphi_{n+2}-\varphi_{n+1}$ and $\psi_{n+2}-\psi_{n+1}$ in terms of $\phi_{n+1}-\phi_n$. To do so, we use extensively $0\le\gamma\le1$, $\kappa\ge0$ and $0\le \phi_n\le1$. In particular, one has $\gamma\phi_n^2-1-\kappa\le0$.

The mean-value theorem gives the bounds
\begin{equation}
\big|\rme^{X_{n+1}}-\rme^{X_n}\big|\le |X_{n+1}-X_{n}|\quad\text{if
$X\le0$},\qquad
|Y_{n+1}^\alpha-Y_n^\alpha|\le\alpha |Y_{n+1}-Y_n|\quad\text{if $|Y|\le1$}.
\label{eq:bd1}
\end{equation}
Furthermore, as can be seen from writing the left-hand side as
$\big|(\rme^{X_{n+1}}-\rme^{X_{n}})Y_{n+1}+\rme^{X_{n}}(Y_{n+1}-Y_n)\big|$,
\begin{equation}
\big|\rme^{X_{n+1}}Y_{n+1}-\rme^{X_n}Y_n\big|\le
|X_{n+1}-X_{n}|+|Y_{n+1}-Y_n|\quad\text{if $X\le0$ and $|Y|\le 1$.}
\label{eq:bd2}
\end{equation}
Then, we write $\varphi_{n+2}-\varphi_{n+1}$ using~\Eq{recur} and bound the different terms arising in this difference using the relations~\eqref{eq:bd1} and~\eqref{eq:bd2}. In practice, we need (for $\alpha=0,1,3$) 
\begin{equation}
\begin{aligned}
&\left|\rme^{\int_u^\tau\rmd
s [ \gamma\phi_{n+1}(s,z+s)^2-1-\kappa]}\phi_{n+1}(u,z+u)^\alpha
 -   \rme^{\int_u^\tau\rmd
s [ \gamma\phi_n(s,z+s)^2-1-\kappa]}\phi_n(u,z+u)^\alpha\right|
\\
&\qquad\qquad\le\int_u^\tau\rmd s\, \gamma \big|\phi_{n+1}(s,z+s)^2-\phi_n(s,z+s)^2\big|
+ \big|\phi_{n+1}(u,z+u)^\alpha-\phi_n(u,z+u)^\alpha\big|
\\
&\qquad\qquad\le 2\gamma \int_u^\tau\rmd s\,
\big|\phi_{n+1}(s,z+s)-\phi_n(s,z+s)\big|
+ \alpha \big|\phi_{n+1}(u,z+u)-\phi_n(u,z+u)\big|
\\&\qquad\qquad\le \gamma\int_u^\tau\rmd
s\,\Delta_n(s)+\tfrac\alpha2\Delta_n(u).
\end{aligned}
\end{equation}
In the last line, we used $|\phi_{n+1}-\phi_n|=\frac12|\varphi_{n+1}+\psi_{n+1}-\varphi_n-\psi_n|\le\frac12\Delta_n$. Then, from~\Eq{recur},
\begin{equation}
\begin{aligned}
&|\varphi_{n+2}(\tau,z+\tau)-\varphi_{n+1}(\tau,z+\tau)|
\le\varphi_0(z)\gamma\int_0^\tau\rmd s\,\Delta_n(s)
\\&\qquad\qquad\qquad\qquad
+\int_0^\tau\rmd u\,\bigg[ (1-\gamma)\Big(\gamma\int_u^\tau\rmd
s\,\Delta_n(s)+\tfrac32\Delta_n(u)\Big)+\kappa\Big(\gamma\int_u^\tau\rmd
s\,\Delta_n(s)+\tfrac12\Delta_n(u)\Big)\bigg].
\end{aligned}
\end{equation}
We already used that $\varphi_0(z)\ge0$. Recalling that $\varphi_0(z)\le1$, we obtain
\begin{equation}
|\varphi_{n+2}(\tau,z+\tau)-\varphi_{n+1}(\tau,z+\tau)|
\le \tfrac a 2 \int_0^\tau \rmd s\,\Delta_n(s)
+\tfrac b 2\int_0^\tau \rmd u\int_u^\tau\rmd s\,\Delta_n(s)
=\tfrac12\int_0^\tau \rmd s\,(a+bs)\Delta_n(s),
\end{equation}
with $a=2\gamma+3(1-\gamma) +\kappa$ and $b=2(1-\gamma+\kappa)\gamma$. The exact same bound also holds, obviously, for $\psi_{n+2}(\tau,z-\tau)-\psi_{n+1}(\tau,z-\tau)|$. Altogether, we finally obtain
\begin{equation}
\Delta_{n+1}(\tau)\le 
\int_0^\tau \rmd s\,(a+bs)\Delta_n(s).
\end{equation}
Obviously, from its definition, $\Delta_0(\tau)\le2$. Furthermore, we have the relation
\begin{equation}
\frac{\big(a \tau +\tfrac12 b \tau^2)^{n+1}}{(n+1)!}
=\int_0^\tau\rmd s\,(a+bs) \frac{\big(a s +\tfrac12 b s^2)^{n}}{n!}.
\end{equation}
Indeed, as functions of $\tau$, both sides have the same derivative and the same value (zero) at $\tau=0$.
Then, by recursion, one obtains
\begin{equation}
\Delta_n(\tau)\le 2 \frac{\big(a \tau +\tfrac12 b \tau^2)^n}{n!}.
\end{equation}
Clearly,~\Eq{eq:goal} holds, and the proof is complete.

Note that, in practice, we have shown that the system~\eqref{systemqq1} admits a solution when the initial conditions satisfy~\Eq{eq:initcondagain}. The same methods can be used to show that the solution is unique.

\section{Profile of the Fermi shock wave} \label{app:Fermi}
In this Appendix, we compute the discontinuity of the Fermi shock wave that develops at the boundary of the light cone and we derive the implicit equation which determines the complete shock wave profile.
We work in the reference frame of the right-moving front by using
\begin{align}
\left\{
\begin{array}{rl}    \displaystyle \phi_+(\tau,z)  :=
\phi(\tau,z+R_0+\tau)&\xrightarrow[\tau\to+\infty]{}f(z)\\[2ex]
    \displaystyle \phi_{1+}(\tau,z)   :=
\phi_1(\tau,z+R_0+\tau)&\xrightarrow[\tau\to+\infty]{}f_1(z),
\end{array}
\right.
\end{align}
where we recall that $R_0$ is the extension of the initial condition. The light cone property implies that, at all times, $\phi_+(\tau,z)=1$ and $\phi_{1+}(\tau,z)=0$ for $z>0$. Then the limiting shapes  also satisfy $f(z)=1$ and $f_1(z)=0$ for $z>0$.
From the coupled PDEs~(5), one obtains the evolution equations for $\phi_+$ and $\phi_{1+}$:
\begin{align}\label{eq:C17} 
\left\{
\begin{array}{rl}
\partial_\tau \phi_+   + \partial_z (\phi_{1+}- \phi_{+})  &=
\phi_+(\phi_{+}^2 - 1) \\[2ex]
\partial_\tau \phi_{1+} - \partial_z (\phi_{1+}-\phi_+)   &=  \phi_{1+} (\gamma  \phi_+^2 - 1-\kappa).
\end{array}
\right.
\end{align}
As $\tau\to\infty$, requiring that $\partial_\tau\phi_+\to0$ and $\partial_\tau\phi_{1+}\to0$ leads to the relations
\begin{align}
 \partial_z (f_1-f) = f(f^2-1)= f_1[1+\kappa-\gamma f^2].
\end{align}
Hence, 
\begin{align}
f_1=f\frac{f^2-1}{1+\kappa-\gamma f^2},
\label{g1g}
\end{align}
and
\begin{align}
 \partial_z \left(f\frac{ f^2-1}{1+\kappa-\gamma f^2}-f\right) = f(f^2-1).
\label{eq:f'f}
\end{align}
Let us now introduce
\begin{equation}
L:=f(0^-),\qquad M:=-f_1(0^-).
\end{equation}
From~\Eq{g1g}, we have
\begin{align}\label{eq:MasL}
    M = L \frac{1-L^2}{1+\kappa - \gamma L^2}.
\end{align}
Now, integrating the first line of~\Eq{eq:C17} over $z\in[-\epsilon,\epsilon]$ yields
\begin{equation}
\partial_\tau\int_{-\epsilon}^\epsilon\rmd
z\,\phi_+(\tau,z)-1-\phi_{1+}(\tau,-\epsilon)+\phi_+(\tau,-\epsilon)=
\int_{-\epsilon}^\epsilon\rmd z\,\phi_+(\tau,z)[\phi_+(\tau,z)^2-1],
\end{equation}
where we used $\phi_{+}(\tau,\epsilon)=1$ and $\phi_{1+}(\tau,\epsilon)=0$. Taking $\tau\to\infty$, and then $\epsilon\to0$ leads to the relation $M+L=1$. Combining with~\Eq{eq:MasL} gives two solutions for $L$ and $M$ (assuming $L>0$):
\begin{align}
\left\{ 
\begin{array}{rl}
    L&= 1  \\[\medskipamount]
    M&= 0
    \end{array}
    \right.
    \mbox{  and  }
    \left\{ 
\begin{array}{rl} \label{eq:solutionsFermi}
    L &= \frac{\sqrt{1+4 \kappa^*(1+\kappa)}-1}{2\kappa^*}\\[\smallskipamount]
    M &= 1-\frac{\sqrt{1+4 \kappa^*(1+\kappa)}-1}{2\kappa^*},
    \end{array}
    \right.
\end{align}
where we recall that $ \kappa^*=\gamma+1$. The expression for $L$ in the second solution is Eq.~(9).

For $\kappa>\kappa^*$, the second solution in~\Eq{eq:solutionsFermi} leads to $L>1$ which cannot be correct, see~\autoref{app:ballistic}.
This implies that one must have $L=1$ and $M=0$: for $\kappa>\kappa^*$, there is no front at late times at position $X\simeq\tau$. As is shown in~\autoref{app:FKPP}, the actual front  propagates at a velocity $v_\kappa<1$ and has all the properties of an FKPP front.

For $\kappa<\kappa^*$, the second solution in~\Eq{eq:solutionsFermi} satisfies $L<1$; this corresponds to a discontinuous profile where $f(z)$ jumps from $L$  to $1$ across the light cone boundary at $z=0$.
Numerically, this is the solution towards which the system converges for $\kappa<\kappa^*$.
We can now solve~\Eq{eq:f'f} for $z<0$ when $\kappa<\kappa^*$, using $f(0^-)=L$ as boundary condition. This is a first-order ordinary differential equation (ODE) that can be integrated by means of decomposition into simple fractions. Implicitly,
\begin{align}
    \frac{2}{1+\kappa- \gamma f^2}
    -\frac{2}{1+\kappa- \gamma L^2}
    +\frac{2+\kappa}{1+\kappa} \log \frac{f^2}{L^2}
    +\frac{ 1 -\kappa + \gamma }{1 +\kappa-\gamma } \log
\frac{1-f^2}{1-L^2}
    -\frac{3+3\kappa-\gamma  }{ (1+\kappa) ( 1+\kappa-\gamma )}\log
\frac{1+\kappa  -    \gamma f^2 }{1+\kappa-\gamma L^2}
  =  2z. \label{eq:implicit}
\end{align}
For $\kappa = 0$ and $\gamma=1$, the solution explicitly reads $L=M=1/2$, $f(z < 0) = 1/\sqrt{1+3\rme^{-z}}$~\cite{ABM}.
As $\kappa\to\kappa^*$ from below, the discontinuity closes continuously as $L\to1$. However, the front does not converge to a flat profile. 
Indeed, linearizing~\Eq{eq:implicit} close to $z = 0^-$ with $f(z < 0) \simeq L + z/\xi$ where $L$ is taken from the second solution in Eq.~(\ref{eq:solutionsFermi}) and $1/\xi := f'(0^-)$, we obtain a finite $\xi(\kappa\to\kappa^*, \kappa<\kappa^*) = 1/4 + \kappa^*/2$.

Exactly at the critical disorder $\kappa = \kappa^*$, the implicit equation reads
\begin{align}
\frac{2}{2+\gamma - \gamma f^2 }
+ \frac{3+\gamma }{2+\gamma}     \left[ \log f^2
- \log \left(2+\gamma - \gamma f^2\right) \right]
= 2 z +1 - \frac{3+\gamma}{2+\gamma} \log 2
\end{align}
and a linearization close to $z = 0^-$ with $f(z < 0) \simeq 1 + z/\xi$ yields $\xi(\kappa = \kappa^*) = 1/2 + \kappa^*$. 
The factor of two between $\xi(\kappa\to\kappa^*, \kappa<\kappa^*) = \lim_{\kappa \to \kappa^{*-}} \lim_{z\to 0^-} 1/f'(z) = 1/4 + \kappa^*/2$ and $\xi(\kappa = \kappa^*) = \lim_{z\to 0^-} \lim_{\kappa \to \kappa^{*-}}  1/f'(z) =  1/2 + \kappa^*$ stems from the existence of a vanishing lengthscale $\propto \kappa-\kappa^*$ which separates two regimes: $f(|z| \ll \kappa - \kappa^* \ll 1)$ and $f( \kappa - \kappa^* \ll |z| \ll 1)$.
However, further analysis or input on the model is required to determine which of the solutions in~\Eq{eq:solutionsFermi} is selected at $\kappa = \kappa^*$.

\section{FKPP dynamics} \label{app:FKPP}
In this Appendix, we discuss how the set of coupled PDEs~(5) can exhibit solutions belonging to the FKPP class~\cite{Fisher1937,KPP1937} for $\kappa > \kappa^*$.
First, in~\autoref{app:hidden_FKPP}, we reformulate these equations to better reveal the backbone of the underlying FKPP physics.
Then, in~\autoref{app:FKPP_prop}, we recall the standard properties of the solutions to an FKPP equation, and we derive the expression Eq.~(12) of the front velocity. This FKPP prediction is tested at $\kappa > \kappa^*$ in Fig.~1 of the main manuscript.
Later, in~\autoref{app:diffu}, we show that the set of coupled PDEs~(5) converges to the actual FKPP equation for $\kappa \gg 1$ (the diffusive limit).
Finally in~\autoref{app:FKPP_check}, we shall show by means of extensive numerical simulations that the solutions to the coupled PDEs~(5) have all the expected properties of an FKPP front as soon as $\kappa>\kappa^*$, and not only in the diffusive limit.

\subsection{Hidden FKPP equation} \label{app:hidden_FKPP}
In this Appendix, we carefully derive Eq.~(11) which results from the coupled PDEs~(5). The latter read
\begin{align}
\left\{     \label{eq:app_coupled_PDEs_adim}
\begin{array}{rl}  
     \partial_\tau \phi +  \nabX \cdot  \boldsymbol{\phi}_1 \!\!\!\!& =  \phi (\phi^2-1)
 \\[\medskipamount]
          \partial_\tau  \boldsymbol{\phi}_1 +  \nabX \phi  \!\!\!\!& = \boldsymbol{\phi}_1  (\gamma\phi^2-1) - \kappa \boldsymbol{\phi}_1.
\end{array}
\right.
\end{align}
Taking the time derivative of the first one and the spatial derivative of the second one, we have
\begin{align}
\left\{    
\begin{array}{rl}  
     \partial^2_\tau \phi +  \nabX \cdot \partial_\tau \boldsymbol{\phi}_1 \!\!\!\!& =   (3\phi^2-1) \partial_\tau \phi
 \\[\medskipamount]
          \nabX \cdot  \partial_\tau  \boldsymbol{\phi}_1 +  \nabX^2 \phi  \!\!\!\!& =  \nabX \cdot  \boldsymbol{\phi}_1  [\gamma\phi^2-1-\kappa]  + 2\gamma  \phi \nabX \phi \cdot \boldsymbol{\phi}_1.
\end{array}
\right.
\end{align}
Using the second equation to eliminate the mixed derivative term in the first one, this yields
\begin{align}
     & \partial^2_\tau \phi -  \nabX^2 \phi  +  \nabX \cdot  \boldsymbol{\phi}_1  [\gamma\phi^2-1-\kappa]  + 2\gamma \phi \nabX \phi \cdot \boldsymbol{\phi}_1 
      =   (3\phi^2-1) \partial_\tau \phi.
\end{align}
Now, $ \nabX \cdot  \boldsymbol{\phi}_1 $ can be expressed in terms of $\phi$ using the first equation in (\ref{eq:app_coupled_PDEs_adim}), yielding 
\begin{align}
     & \partial^2_\tau \phi -  \nabX^2 \phi  + [- \partial_\tau \phi + \phi(\phi^2-1) ]
     [\gamma\phi^2-1-\kappa] + 2\gamma  \phi\nabX \phi \cdot \boldsymbol{\phi}_1  =   (3\phi^2-1) \partial_\tau \phi.
\end{align}
We switch to the parameter $\kappa^* := 1+\gamma$ and reorganize to only have on the {\sc lhs} derivative terms with a constant prefactor.
\begin{equation}\label{eq:app_coupled_PDEs1} 
\left\{    
\begin{array}{rl}  
   \partial_\tau^2 \phi +(\kappa-\kappa^*) \partial_\tau \phi - \nabX^2 \phi &=   
   \phi (\phi ^2-1) [  (2+\kappa-\kappa^*) + (1-\kappa^*) (\phi ^2 -1) ]  \\
   & \qquad + (2+\kappa^*) (\phi^2-1) \partial_\tau \phi  - 2 (\kappa^*-1) \phi \boldsymbol{\phi}_1 \cdot \nabX \phi \\[\smallskipamount]
   \partial_\tau \boldsymbol{\phi}_1 + \nabX \phi &= \boldsymbol{\phi}_1 [ (\kappa^*-1)\phi^2-1-\kappa ].
   \end{array}
   \right.
\end{equation}
The derivative terms on a {\sc rhs} cancel when $\phi\to 1$ and $\boldsymbol{\phi}_1\to 0$.
Finally, we switch to the field $\delta\phi := 1-\phi$ and  obtain the following set of coupled PDEs
\begin{align} 
\left\{    \label{eq:FKPP_app_0}
\begin{array}{rl}  
   \partial_\tau^2 \delta\phi + (\kappa-\kappa^*) \partial_\tau \delta\phi - \nabX^2 \delta\phi &=   
   \delta\phi (1-\delta\phi) (2-\delta\phi) [ (2+\kappa-\kappa^*) - (1-\kappa^*) \delta\phi(2-\delta\phi) ] \\
   &\qquad - (2+\kappa^*) \delta\phi (2-\delta\phi) \partial_\tau \delta\phi  - 2 (\kappa^*-1) (1-\delta\phi) \boldsymbol{\phi}_1 \cdot \nabX \delta\phi \\[\smallskipamount]
   \partial_\tau \boldsymbol{\phi}_1 - \nabX \delta\phi &=-( 2+\kappa-\kappa^*) \boldsymbol{\phi}_1 - (\kappa^*-1) \boldsymbol{\phi}_1 \delta\phi (2-\delta\phi) .
   \end{array}
   \right.
\end{align}
We now re-express the {\sc rhs} of both equations above by collecting the non-linear terms under the symbols $\mathrm{NL}$ and $\mathrm{NL}_1$. 
This yields
\begin{align} \label{eq:FKPP_app}
\left\{    
\begin{array}{rl}  
   \partial_\tau^2 \delta\phi + (\kappa-\kappa^*) \partial_\tau \delta\phi
- \nabX^2 \delta\phi &= 2(2+\kappa-\kappa^*) \delta\phi - \mathrm{NL}\\[\smallskipamount]
   \partial_\tau \boldsymbol{\phi}_1 - \nabX \delta\phi &=-( 2+\kappa-\kappa^*) \boldsymbol{\phi}_1 - \mathrm{NL}_1 .
   \end{array}
   \right.
\end{align}
The initial conditions are $\delta\phi(\tau=0,\bX) = \delta\phi_0(X)$, $\boldsymbol{\phi}_1(\tau=0,\bX) = 0$, and  $\partial_\tau \delta\phi(\tau=0,\bX) \simeq 2 \delta\phi_0(X)$.
Notice the dependence on $\boldsymbol{\phi}_1$ in the first equation which has not been fully eliminated but is now entering at a non-linear level. Naturally, one may check that $\textsc{rhs}$ vanishes at the correlated- and uncorrelated-world solutions given by ($\delta\phi=\boldsymbol{\phi}_1=0$) and ($\delta\phi=1$, $\boldsymbol{\phi}_1=0$), respectively. This concludes the derivation of Eq.~(11).

\subsection{Properties of the FKPP solutions -- Short introduction to FKKP-ology} \label{app:FKPP_prop}
The FKPP equation was introduced independently in 1937 by Fisher~\cite{Fisher1937} and Kolmogorov-Petrouvsky-Piscounov~\cite{KPP1937} to describe problems in biology related to evolution. It has been the subject of many works and studies since. An extensive review was published in 2003~\cite{vanSaarloos.2003,BrunetHDR}.
We only consider the one-dimensional $d=1$ case, and focus on $\phi(\tau,X)$ only, regarding $\phi_1$ as an intermediate calculation step. We begin by giving a list of defining characteristics that an evolution equation for $\phi(\tau,X)$ should satisfy in order to belong to the FKPP class. By evolution equation, here we mean the actual equation for times $\tau>0$, without the input of the specific initial condition at time $\tau = 0$.
\begin{enumerate} 
\item \emph{The equation admits a stable uniform solution and an unstable uniform solution.}\\ 
In our case, $\phi=1$ is an unstable solution (also called the unstable phase), while $\phi=0$ is a stable solution. (In both phases, we also need to take $\phi_1=0$ but, as already mentioned, we focus on $\phi$ only.)
\item \emph{The equation admits traveling wave solutions for all velocities 
larger or equal than a critical velocity $v^*$.}\\ 
This means that, for all $v\ge v^*$, there exists functions $f_v(z)$ with $f_v(\infty)=1$ and $f_v(-\infty)=0$ such that $f_v(X-v\tau)$ is a solution. Depending on the initial condition, the front $\phi(\tau,X)$ properly centered might converge to any of these traveling waves.
\item \emph{The traveling wave solutions decay exponentially towards the unstable phase~1, with a linear prefactor for the critical traveling wave only.}\\ 
This means that, there is a spatial decay rate $\mu=\mu(v)>0$ and $\mu^*:=\mu(v^*)$ such that
\begin{equation}
1-f_v(z) \propto  \exp({-\mu z}) \text{ as } z\to\infty \text{ for } v>v^*,\quad \text{  and }
1-f_{v^*}(z) \simeq A \, z  \exp({-{\mu^*} z}) \text{ as
} z\to\infty.
\label{Atw}
\end{equation}
\textit{Nota}: in the main manuscript, we wrote $\mu$ and $v$ instead $\mu^*$ and $v^*$ as there was no possible ambiguity. The traveling waves are defined up to translation: if $f_{v^*}(z)$ is a traveling wave at velocity $v^*$, then so is $f_{v^*}(z-\delta)$. In this Appendix, we choose the particular critical traveling wave such that $f_{v^*}(z)=1-[A\, z+\epsilon(z)]\exp(-\mu^* z)$, where $\epsilon(z)\to0$ as $z\to\infty$. The quantity $A$ can be measured numerically.

\item \emph{When $\delta\phi_0(X) := 1-\phi(0,X)$ decays faster than $\exp({-\mu^* X})$ for large $X$, then the front $\phi(\tau,X)$ propagates asymptotically at velocity $v^*$ and its shape converges to the critical traveling wave.}\\
If we call $m_\tau$ the position of the front at time $\tau$, this means that $m_\tau\sim v^*\tau$ as $\tau\to\infty$ and that
\begin{equation}
\phi(\tau, m_\tau +\delta+z) \to f_{v^*}(z)\quad\text{as $\tau\to\infty$,}
\label{Aphif}
\end{equation}
where $\delta$ is a constant offset depending on the initial condition and on the precise definition of $m_\tau$.

\item \emph{Still assuming that $\delta\phi_0(X) := 1-\phi(0,X)$ decays faster than $\exp({-\mu^* X})$ for large $X$, the asymptotic position of the front is}
\begin{equation}
m_\tau=v^*\tau -B \log \tau +a -\frac{E}{\sqrt\tau}+\mathcal
O\Big(\frac{\log\tau}\tau\Big)\quad\text{as $\tau\to\infty$}.
\label{m_t}
\end{equation}
The quantities $B$ and $E$ are known (see below) and independent of the initial condition.
The terms $-B\log\tau$ and $-E/\sqrt\tau$ are respectively known as the Bramson term~\cite{Bramson.1978,Bramson.1983} (see also Refs.~~\cite{HamelNolenRoquejoffreRyzhik.2013,Roberts.2013}) and the Ebert-van Saarloos term~\cite{EbertvanSaarloos.2000} (see also Refs.~\cite{BBHR.2016,NolenRoquejoffreRyzhik.2019}). The next term, of order $(\log\tau)/\tau$, has also a known coefficient~\cite{BerestyckiBrunetDerrida.2017,BerestyckiBrunetDerrida.2018,Graham.2019}. The constant offset $a$ depends on the initial condition and on the precise definitions of $m_\tau$.

\item \emph{Still assuming that $\delta\phi_0(X) := 1-\phi(0,X)$ decays faster than $\exp({-\mu^* X})$ for large $X$, one has for the shape of the front:}
\begin{equation}
1-\phi(\tau,m_\tau+\delta+z)\simeq A\, z \exp\left(-\mu^* z- C \frac{z^2}\tau\right)\qquad\text{as $\tau\to\infty$ and $z\propto\sqrt\tau$},
\label{gauss}
\end{equation}
where $A$ and $\delta$ are the same as in~\eqref{Atw} and~\eqref{Aphif}.
The quantity $C$ can be computed (see below).
\end{enumerate}
In our problem, the initial conditions $\delta \phi_0(X)$ are defined on a finite support of radius $R_0$ and thus do decay faster than $\exp({-\mu^* X})$ for large $X$. This implies that points 4., 5., and 6. above do apply. In practice, in order to identify $v^*$,  the first step is to look for solutions satisfying
\begin{equation}\label{Adecay}
1-\phi(\tau,X)=f_v(X-v\tau)\propto \exp\left(-\mu (X-v\tau)\right)\quad\text{for large $z:=X-v\tau$}.
\end{equation}
For that purpose, we can consider the equation linearized around $\phi=1$, in the unstable phase. This leads to a relation between $v$ and $\mu$, which we can write $v=v(\mu)$. A necessary condition to have an FKPP equation (see point 2.\@ above) is that $v(\mu)$ reaches a minimum when $\mu=\mu^*$ for some finite value $\mu^*$:
\begin{equation}\label{Av*}
v^*:=\min_\mu v(\mu) = v(\mu^*).
\end{equation}

For the problem at hand, we inject~\Eq{Adecay} into~\Eq{eq:FKPP_app}, where we recall that $\delta\phi:=1-\phi$. 
Neglecting non-linear terms, this leads to
\begin{equation}
(\mu v)^2+(\kappa-\kappa^*)\mu v -\mu^2=2(2+\kappa-\kappa^*).
\end{equation}
The solution with $v>0$ is
\begin{equation}\label{AVmu}
v(\mu)=\frac{\sqrt{(4+\kappa-\kappa^*)^2+4\mu^2}-(\kappa-\kappa^*)}{2\mu}.
\end{equation}
For $\kappa>\kappa^*$, one can see that
$v(\mu)$ reaches a minimum at a finite $\mu^*$.
Conversely, for $\kappa\le\kappa^*$, $v(\mu)$ is always decreasing and, thus, does not reach a minimum. This indicates that the problem cannot be FKPP if $\kappa\le\kappa^*$, and that it might be FKPP if $\kappa>\kappa^*$.

From now on, let us assume that $\kappa>\kappa^*$.
A simple computation from~\Eq{AVmu} yields the selected front velocity and spatial decay rate
\begin{equation}
v^*= \frac{2\sqrt{4+2(\kappa-\kappa^*)}}{4+\kappa-\kappa^*},
\qquad
\mu^* =\frac{4+\kappa-\kappa^*}{\kappa-\kappa^*}\sqrt{4+2(\kappa-\kappa^*)}.
\label{eq:mu*v*}
\end{equation}
This is the solution announced in Eq.~(12) of the main manuscript.
For an FKPP front, the behavior of $v(\mu)$ around $\mu=\mu^*$ controls many properties of the propagating front. In particular, the quantities $B$ and $E$ appearing in~\eqref{m_t}, and the quantity $C$ appearing in~\eqref{gauss} are given by
\begin{equation}
B=\frac3{2\mu^*},\qquad
E=3\sqrt{\frac{2\pi}{\mu^{*5} \, v''(\mu^*)}},\qquad
C=\frac1{2\mu^*v''(\mu^*)}.
\label{Apredictvalues}
\end{equation}
See for instance Ref.~\cite{Bramson.1983} for $B$, and Ref.~\cite{EbertvanSaarloos.2000} for $E$. The above expression for $C$ can be obtained with the methods developed in Ref.~\cite{BrunetDerrida.1997}. 

\subsection{FKPP equation in the diffusive regime, \texorpdfstring{$\kappa \gg1$}{κ≫1}.}
\label{app:diffu}
Here, we derive the asymptotic FKPP theory when the disorder strength is very large, $\kappa \gg 1$.
We start from the re-formulation of the coupled PDEs in~\Eqs{eq:app_coupled_PDEs1} which, we recall, read
\begin{align} 
\left\{    \label{eq:app_coupled_PDEs}
\begin{array}{rl}  
   \partial_\tau^2 \phi +(\kappa-\kappa^*) \partial_\tau \phi - \nabX^2 \phi &=   
   \phi (\phi ^2-1) [  (2+\kappa-\kappa^*) + (1-\kappa^*) (\phi ^2 -1) ] \nonumber \\
   & \qquad + (2+\kappa^*) (\phi^2-1) \partial_\tau \phi  - 2 (\kappa^*-1) \phi \boldsymbol{\phi}_1 \cdot \nabX \phi \\[\smallskipamount]
   \partial_\tau \boldsymbol{\phi}_1 + \nabX \phi &= \boldsymbol{\phi}_1 [ (\kappa^*-1)\phi^2-1-\kappa ].
   \end{array}
   \right.
\end{align}
When $\kappa$ is large, the exponential spatial decay rate $\mu^*$ is of order $\sqrt\kappa$, see~\Eq{eq:mu*v*}. This indicates that the width of the front is small, of order $1/\sqrt\kappa$. This suggests rescaling space by working with
\begin{equation}
\bY:=\sqrt{2\kappa} \bX.
\end{equation}
Keeping only the terms of order $\kappa$, we see that the rescaled equation for $\phi$ converges as $\kappa\to\infty$ to
\begin{equation}
\partial_\tau\phi-2 \nabY^2\phi=\phi(\phi^2-1),
\label{FKPPY}
\end{equation}
which is the FKPP equation.
Notice that the dependence on $\gamma$, the distance to the superconducting transition, has dropped.
A standard analysis of the above FKPP equation, following the steps in~\autoref{app:FKPP_prop}, yields a front traveling at the velocity $4$ on the $\bY$ scale, and thus with a velocity of order $2\sqrt{2/\kappa}$ on the $\bX$ scale, in agreement with~\Eq{eq:mu*v*}.
The radial profile of the critical traveling wave corresponding to
\Eq{FKPPY} must satisfy the ODE
\begin{equation}\label{eq:ODE_FKPP}
2\hat  f''+ 4 \hat f' + \hat f(\hat f^2-1) = 0.
\end{equation}
The prediction is that the late-time shape of the front, $f_{v^*}(z)$ defined in~\Eq{Aphif}, converges after rescaling to
\begin{equation}\label{eq:FKPP_scale}
f_{v^*}(Y/\sqrt{2\kappa})\to \hat f(Y)\qquad\text{as $\kappa\to\infty$}.
\end{equation}
This is numerically confirmed in Fig.~3 of the main manuscript.
(\textit{Nota}: the solutions to~\Eq{eq:ODE_FKPP} are defined up to translation.)

\subsection{Numerical evidence of FKPP dynamics at 
\texorpdfstring{$\kappa > \kappa^*$}{κ>κ*}  -- FKPP-ometry} \label{app:FKPP_check}
Here, we present numerical simulations in $d=1$ that demonstrate that $\phi(\tau,X)$ behaves as an FKPP front as soon as $\kappa>\kappa^*$, and not only when $\kappa\gg\kappa^*$ where we showed in~\autoref{app:diffu} that the equations for $\phi$ were converging to the actual FKPP equation. Specifically, we show that although the equations for $\phi$ look quite different from the FKPP equation at finite $\kappa>\kappa^*$, the FKPP properties listed in~\Eqs{Aphif},~\eqref{m_t} and~\eqref{gauss} hold. This is in addition to the numerical results presented in Fig.~1 of the main manuscript where we tested the FKPP prediction for the front velocity when $\kappa > \kappa^*$.

\subsubsection{Discretization and integration schemes}
Checking that the late-time~\Eq{m_t} holds requires a very accurate prediction for the velocity of the front. However, as shown below, the discretization scheme of the equations as well as their integration scheme slightly impact the value of that velocity. This implies that the analytical prediction for the velocity must take into account the specific numerical integration scheme that is used.
Another difficulty is that FKPP fronts are extremely sensitive to the exponential tail ahead of the front, at a distance on the order of $\sqrt\tau$ ahead of the front, where $\phi(\tau,X)$ is very close to 1. This requires a numerical solver that remains very accurate in that region.
For these reasons, we found it easier to write our own solver, which we now describe.

Taking the sum and difference of the two coupled PDEs~(5), we obtain
\begin{equation}
\partial_\tau(\phi\pm\phi_1)
\pm\partial_X(\phi\pm\phi_1)= g_\pm(\phi,\phi_1).
\end{equation}
with the functions $g_\pm(\phi,\phi_1) := \phi(\phi^2-1) \pm \phi_1[\gamma\phi^2-1-\kappa]$.
This is equivalent to the following set of coupled first-order ODEs,
\begin{equation}
\frac{\rmd}{\rmd\tau}\Big[
(\phi\pm\phi_1)(\tau,\pm\tau+z)\Big]
=g_\pm\left(\phi(\tau,\pm\tau+z),\phi_1(\tau,\pm\tau+z)\right).
\end{equation}
We discretize them with a simple Euler scheme, using a common discretization step $a$ for both space and time. This reads
\begin{equation}
(\phi\pm\phi_1)(\tau+a,\pm(\tau+a)+z)=
(\phi\pm\phi_1)(\tau  ,\pm\tau  +z)+a
g_\pm\left(\phi(\tau,\pm\tau+z),\phi_1(\tau,\pm\tau+z)\right),
\end{equation}
or, writing $X=\pm(\tau+a)+z$,
\begin{equation}
(\phi\pm\phi_1)(\tau+a,X)=
(\phi\pm\phi_1)(\tau  ,X\mp a)+a
g_\pm\left(\phi(\tau,X\mp a),\phi_1(\tau,X\mp a)\right),
\end{equation}
Finally, solving for $\phi$ and $\phi_1$,
\begin{align}
\hspace{-0.5em}
\left\{
\begin{array}{rl}
\phi(\tau+a,X)=& \frac{\phi(\tau,X-a)+\phi(\tau,X+a)}2
+\frac{\phi_1(\tau,X-a)-\phi_1(\tau,X+a)}2
+a\frac{g_+\left(\phi(\tau,X-a),\phi_1(\tau,X-a)\right)+g_-\left(\phi(\tau,X+ a),\phi_1(\tau,X+ a)\right)}2\\[\medskipamount]
\phi_1(\tau+a,X)=&\frac{\phi(\tau,X-a)-\phi(\tau,X+a)}2
+\frac{\phi_1(\tau,X-a)+\phi_1(\tau,X+a)}2
+a\frac{g_+\left(\phi(\tau,X-a),\phi_1(\tau,X-a)\right)-g_-\left(\phi(\tau,X+ a),\phi_1(\tau,X+ a)\right)}2.
\end{array}    
\right. \label{phidisc}
\end{align}
These equations allow efficient numerical computation of $\phi(na,pa)$ for any $n\in\mathbb N$ and $p\in\mathbb Z$ as a function of the discretized initial condition $\phi(0,pa)$ for $p\in\mathbb Z$.
Let us make a few practical remarks:
\begin{itemize}
\item In this scheme, the discretization step $a$ is the same for space and time. This guarantees a strict light-cone structure: if the initial condition is zero for $X>R_0$, then $\phi(\tau,X)=0$ for $X> R_0+\tau$.
\item At a given time step, the values on even lattice sites only depend on the values on odd lattice sites at the previous time step, and \textit{vice versa}. In other words, the even and odd lattice sites at a given time step are independent realizations with slightly different initial conditions. Unless the initial conditions are extremely fine-tuned, this leads to a front shape that is not smooth. To avoid this issue, we only display even lattice sites at even time steps. 
\item FKPP equations are very sensitive to what happens in the region where the front is close to the unstable phase, here $\phi=1$. As the default representation of reals in a computer (we use the ``long double'' type) is much better at handling numbers close to 0 rather than numbers close to 1, we simulate $\delta\phi:=1-\phi$ rather than $\phi$ directly. 
\item To save computation time and to avoid instabilities due to loss of precision for numbers too close to 0 or 1, we truncate values at each time step, setting to 0 any value of $\delta\phi$ smaller than $10^{-2000}$, and setting to 1 any value larger than $1-10^{-16}$. 
\end{itemize}

The discretized equations~\eqref{phidisc} lead to a front that propagates at a velocity slightly different from the one predicted by the continuous equations. In fact, we can redo the analysis of~\autoref{app:FKPP_prop}, looking for a traveling wave solution with an exponential decay $1-\phi(\tau,X)\propto \rme^{-\mu(X-v\tau)}$. This predicts a relation
\begin{equation}
v(\mu)=\frac{\log\cosh(\mu a)}{\mu a}+\frac1{\mu
a}\log\!\left[\!1+
\frac{a}{2} \left( \sqrt{(4+\kappa-\kappa^*)^2
+4\frac{\tanh(\mu
a)^2}{a^2}\Big(1- a(\kappa-\kappa^*)-2a^2(2+\kappa-\kappa^*)\Big)} \!-\!  (\kappa-\kappa^*) \right)
\right],
\label{AVmua}
\end{equation}
which reduces to~\Eq{AVmu} when $a\to0$.

\subsubsection{Numerical tests of the FKPP predictions}
We computed $\phi(\tau,X)$ for $\gamma=1$ (\textit{i.e.} $\kappa^*=2$) and $\kappa=3$ up to times $\tau = 10^5$ starting from the $C^1$ initial condition 
\begin{align} \label{eq:initial_cond_numerics}
   \delta \phi_0(X) := 1-  \phi(0,X)=0.2 (1-X^2)^2\, \Theta(1-|X|),
\end{align}
 where $\Theta(X)$ is the Heaviside step function.
We took a discretization step $a=10^{-3}$ for which the minimum of~\Eq{AVmua} predicts
\begin{equation}
v^*\approx 0.979\,551\,163,\qquad
\mu^*\approx 12.1774,\qquad
v''(\mu^*)\approx 2.674\,10^{-4}.
\label{Aval}
\end{equation}
These predicted values are to be compared with $v^*=\frac25\sqrt6\approx 0.979\,796$, $\mu^*=5\sqrt6\approx 12.2474$ and $v''(\mu^*)=5^{-5}\sqrt{2/3}\approx 2.613\,10^{-4}$ obtained from the continuous equations, see~\Eq{eq:mu*v*}.

In practice, we locate the front using the following definition:
\begin{equation}\label{mtau}
m_\tau:= 2a \sum_{n\ge0} [1-\phi(\tau,2an)]\simeq \int_0^\infty\rmd
X \,[1-\phi(\tau,X)].
\end{equation}
This definition of the position only takes into account even lattice sites, as per the discussion above. As can be seen in~\autoref{shape1} below, it turns out that~\Eq{mtau} is nearly equivalent to defining $m_\tau$ as the position where the front is equal to 0.6.

In~\autoref{shape1}, we plot the shape of the front at early times (up to $\tau=10$), and we observe that, to the eye, it converges quite quickly to the limiting shape, as per point 4.\@ of~\autoref{app:FKPP_prop}. The velocity of the front seems already to be smaller than 1.
In~\autoref{shape2}, we plot the shape of the front at larger times (up to  $\tau=10^5$), appropriately rescaled such as to check point 6.\@ of~\autoref{app:FKPP_prop}. Finally, in~\autoref{figposition}, we compare $m_\tau$ obtained from the program with the asymptotic prediction~\eqref{m_t} given in point 5.

The data in these three figures unambiguously demonstrate that the function $\phi(\tau,X)$ has all the properties expected for an FKPP front.

\begin{figure}[!htp]
\centering
\includegraphics[width=110mm]{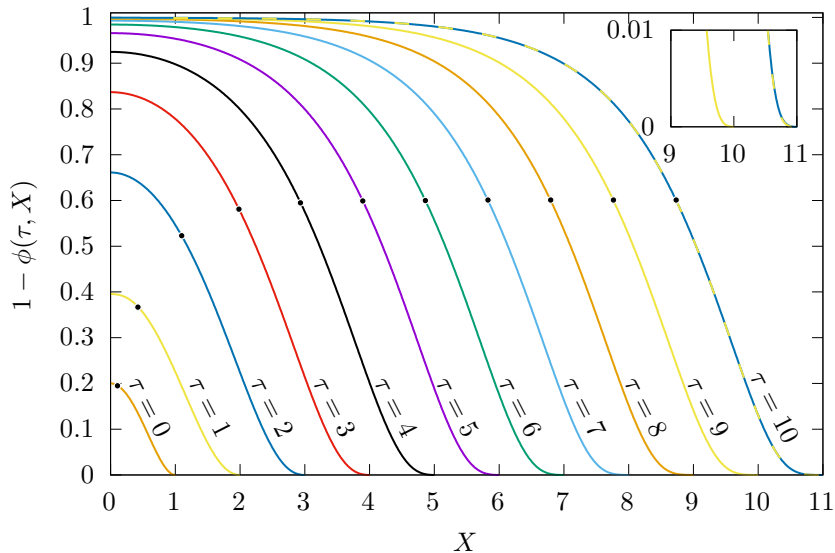}
\caption{$\delta\phi(\tau,X) := 1-\phi(\tau,X)$ as a function of $X$, for $\tau=0,1,\ldots,10$. 
The initial condition at $\tau = 0$ is given in~\Eq{eq:initial_cond_numerics} and extends to $R_0 = 1$.
On each curve, we placed a black dot  at the coordinate corresponding to the front position $m_\tau$ defined in~\Eq{mtau}. To test the convergence of the overall front shape at $\tau=10$, we superimposed in a dashed line the (yellow) curve for $\tau=9$ shifted by $m_{10}-m_9\simeq0.96945$. To the eye, the shape of the front has already converged. The quantity $m_{10}-m_9$ gives an estimate for the velocity which is smaller than the predicted velocity $v^*$ from~\Eq{Aval}; this is due to the sublinear terms in~\Eq{m_t}.
The curve for a given $\tau$ stops at $X=\tau+R_0$ (and is implicitly equal to 0 after that point). This is a direct consequence of the light cone property. The inset shows a zoom of the bottom-right corner of the main plot. ($\gamma = 1$, $\kappa=3$.)
}
\label{shape1}
\end{figure}

\begin{figure}[!htp]
\centering
\includegraphics[width=110mm]{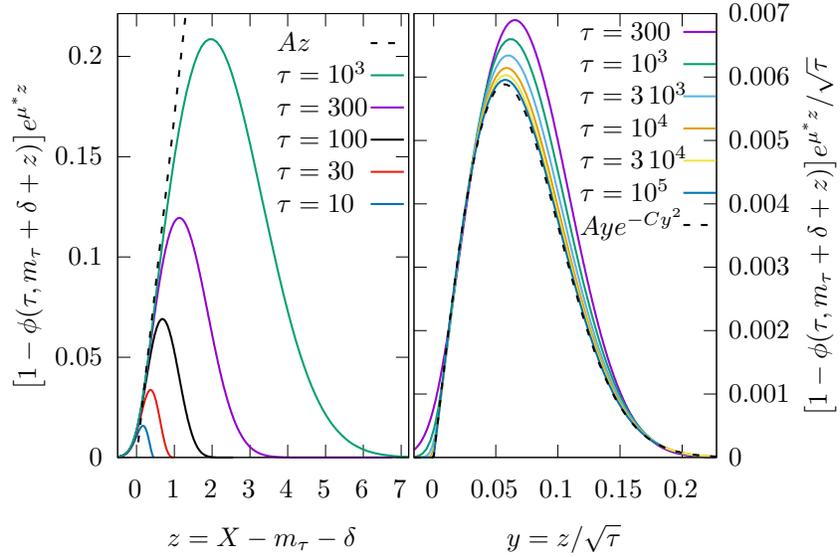}
\caption{
Test of the prediction~\eqref{gauss} for the shape of the front.
(left) $\big[1-\phi(\tau,m_\tau+\delta+z)]\rme^{\mu^* z}$ as a function of $z=X- m_\tau-\delta$, \textit{i.e.} the fronts are centered around their positions and the predicted exponential decay has been expunged in order to study $\delta\phi:=1-\phi$ in the region where it is small. 
(right) The same quantity with both axis now rescaled by $\sqrt\tau$ converges to $Ay\exp({-C y^2})$ (dashed line) as $\tau\to\infty$. The values of $\delta$ and $A$ were adjusted manually and the value of $C$ was computed from~\Eq{Apredictvalues}.
($\gamma = 1$, $\kappa=3$.)
}
\label{shape2}
\end{figure}

\begin{figure}[!htp]
\centering
\includegraphics[width=110mm]{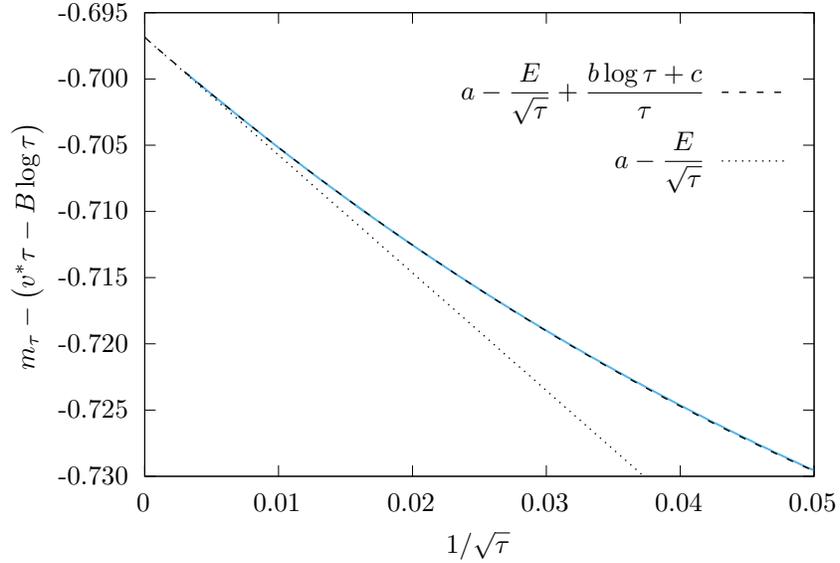}
\caption{Test of the prediction~\eqref{m_t} for the position of the front as a function of time.
In blue, we plot the position $m_\tau$ minus the first two terms of the expansion~\eqref{m_t}, which are known, as a function of $1/\sqrt\tau$, for $\tau$ up to $10^5$.
The result should asymptotically be given by $a-E/\sqrt\tau$, where $E$ is known, but not $a$.
The data is compared to the function $a-E/\sqrt\tau +(b\log\tau+c)/\tau$ (dashed line), using the known value of $E$ and fitting for $a$, $b$ and $c$. The dotted line is the simpler asymptotic expansion $a-E/\sqrt\tau$. To illustrate the precision of this test, the position of the front is equal to $m_\tau\simeq9793.6719$ at $\tau=10^4$; the plotted value is $m_\tau-(v^*\tau-B\log\tau)\simeq-0.7052$ while $a-E/\sqrt\tau\simeq-0.7058$. Notably, all the digits of the value of $v^*$ given in~\Eq{Aval} are needed to produce this figure. ($\gamma = 1$, $\kappa=3$.)
}
\label{figposition}
\end{figure}

%\bibliography{main.bib}

%\end{document}

\end{document}